\documentclass[]{bytedance_seed}

\usepackage[toc,page,header]{appendix}

\usepackage{minitoc}
\newcommand{\sys}{\text{MegaScale-Data}}

\usepackage{tikz}
\newcommand{\circlednumber}[1]{%
    \tikz[baseline=(char.base)]{
        \node[shape=circle,draw,fill=black,text=white,inner sep=1pt] (char) {#1};
    }
}

\usepackage{mathtools}
\DeclarePairedDelimiter\ceil{\lceil}{\rceil}

\usepackage{algorithm}
\usepackage{algpseudocode}
\usepackage{amsmath}

\title{\sys{}: Scaling Dataloader for Multisource Large Foundation Model Training }

\author[1,2,\circ]{Juntao Zhao}
\author[1,\circ]{Qi Lu}
\author[1,\circ,\dagger]{Wei Jia}
\author[1,2]{Borui Wan}
\author[1]{Lei Zuo}
\author[1]{Junda Feng}
\author[1]{Jianyu Jiang}
\author[1]{Yangrui Chen}
\author[1]{Shuaishuai Cao}
\author[1]{Jialing He}
\author[1]{Kaihua Jiang}
\author[1]{Yuanzhe Hu}
\author[1]{Shibiao Nong}
\author[1,\dagger]{Yanghua Peng}
\author[1,\dagger]{Haibin Lin}
\author[2,\dagger]{Chuan Wu}

\affiliation[1]{ByteDance Seed}
\affiliation[2]{The University of Hong Kong}

\contribution[\circ]{Equal Contribution}
\contribution[\dagger]{Corresponding authors}

\abstract{
Modern frameworks for training large foundation models (LFMs) employ dataloaders in a data-parallel manner, with each loader processing a disjoint subset of training data. 
When preparing data for LFM training that originates from multiple, distinct sources, two fundamental challenges arise. First, due to the quadratic computational complexity of the attention operator, the non-uniform sample distribution over data-parallel ranks leads to significant workload imbalance among dataloaders, degrading the training efficiency. Second, supporting diverse data sources requires per-dataset file access states that are redundantly replicated across parallel loaders, consuming excessive memory. This also hinders dynamic data mixing (e.g., curriculum learning) and causes redundant access/memory overhead in hybrid parallelism.

We present \sys{}, an industrial-grade distributed data loading architecture for multisource LFMs training, with three key innovations: (1) Disaggregated data preprocessing via role-specific actors (Source Loaders/Data Constructors) to eliminate source and parallelism redundant data access and ensure multisource scalability.
(2) Centralized and declarative data plane for load-time multisource orchestration, such as long-short context, multimodality, and curriculum learning.
(3) Multi-level auto-partitioning and scaling mechanism for source loaders under heterogeneous preprocessing costs. We also contribute our designs and operational experience in deployment and fault tolerance. \sys{} achieves up to:
(1) 4.5$\times$ end-to-end training throughput improvement, and (2) 13.5$\times$ reduction in CPU memory usage. 
}

\date{\today}

\begin{document}
\maketitle


\section{Introduction}
The rise of large language and vision models, aka large foundation models (LFM), has propelled numerous downstream applications to achieve remarkable performance. Training LFMs faces 
major challenges on model efficiency and data efficiency. Model efficiency 
is related to effectively distributing a massive model 
parameters among GPUs to maximize resource utilization and reduce resource waste. Significant advances in training framework design~\cite{megatron, megatron_lm_train, torch_fsdp, rasley2020deepspeed, jiang2024megascale} and novel parallelism paradigms~\cite{ringattention, infiLLM, ds_ulysses} have been made, largely addressing model efficiency. Data efficiency, on the other hand, ensures that sufficient 
resources (e.g., CPU cores and DRAM) are allocated to the data preprocessing pipeline to avoid input data stalls during training and underutilization of valuable GPU hours. 

Modern training frameworks partition the global training batch across data-parallel dataloaders, with each loader tasked with loading and preprocessing its assigned data subset. Specifically, LFM training data stems from diverse multilingual and multimodal sources, such as text (Wikipedia, Common Crawl~\cite{commoncrawl}), images (ImageNet~\cite{imagenet}, LAION~\cite{laion}), and domain-specific collections~\cite{llama3}. To meet the needs of LFM training, engineers typically depend on ad-hoc scripts to blend these datasets into \textit{data mixtures}~\cite{gemini, llama3, ds_llm}, which combine data sources at specified ratios for different use cases. Although some prior works~\cite{fastflow, graur2022cachew, pecan, tf_data_service, tf.data, DPP} have delved into data efficiency in model training, multisource preprocessing in LFM introduces unique challenges.


\noindent\textbf{Multisource Data Orchestration.}
Under non-uniform sample distributions from diverse sources, the quadratic computational complexity of the attention operator~\cite{vaswani2023attentionneed} introduces significant workload imbalances. For example, a complete sequence composed of 30-token and 70-token subsequences incurs 16\% more computation than two 50-token subsequences. These imbalances are manifested both intra- and inter-microbatch~\cite{DistTrain}, creating stragglers over data parallel ranks, exacerbating pipeline bubbles over pipeline stages, and prolonging training time. The data imbalance 
is compounded by non-uniform data distribution across modality modules. In vision-language models (VLMs)~\cite{qwen2_vl, ds_vl2, janus}, for instance, image encoders process raw pixels while language backbones operate on fused image-patch and text tokens. In that way, modality data within a training batch renders different workloads across modules within the VLM. 

\noindent
\textbf{Multisource Scalability}. Constructing data mixtures from massive sources introduces fundamental memory constraints. 
{\em First}, each dataloader worker 
process maintains independent data file access states via dedicated per-source resources: separate socket connections, schemas, metadata structures, and I/O buffers (e.g., Parquet Row Group~\cite{parquet_apache}). This architecture imposes linear memory overhead growth concerning the number of data sources—a critical limitation given that modern LFM training incorporates hundreds of data sources. {\em Second}, practitioners perform multimodal \textit{transformations} (JPEG/RGB conversion, video keyframe extraction~\cite{keyframe,pytorch_video}) at runtime to avoid inflated storage (up to 200$\times$ for OCR). Such heavyweight processing necessitates vertical scaling of worker numbers to prevent data stalls, and dramatically increases the memory pressure of the dataloader.
Disparities across data sources exacerbate the situation. 
For instance, audio processing requires 4$\times$ more computation per output token than image decoding and 300$\times$ more than text tokenization. Analogous heterogeneity also occurs in unimodal contexts through variable resolutions in images/videos. This forces the loader worker number to be sized for the data source with the largest preprocessing cost, creating severe resource over-provisioning.


Further, the \textit{data mixture} is often dynamic. For instance, curriculum learning strategies~\cite{soviany2022curriculumlearningsurvey} require the data mixture to evolve during training, starting with conceptually "easier" samples and progressively increasing the proportion of more challenging data. Similarly, other methods adapt the data mix based on real-time training metrics like loss~\cite{ADO}. This demands a data preprocessing framework that can express arbitrary mixing schedules and scale efficiently with evolving dataset preprocessing costs. The complexity is compounded by the hybrid parallelism strategies essential for LFM training. In common setups like pipeline parallelism (PP) or context parallelism (CP), multiple GPUs (ranks) collaborate on the same batch of data. For example, in a pipeline, different GPUs form stages that process a batch sequentially; in context parallelism, different GPUs process different segments of the same long sequence. Without a coordinated delivery system, the default approach is for each GPU's process to run a separate, identical dataloader. This naive approach causes massive waste, as the same data is redundantly fetched, preprocessed, and stored in memory on multiple devices, consuming significant I/O bandwidth and memory.

Current dataloaders~\cite{torch_dataloader, tf.data, tf_data_service, ray_data, graur2022cachew, pecan} are ill-equipped for the complexities of LFM training, as they are primarily designed for \textbf{single-source, data-parallel} scenarios rather than \textbf{multisource, hybrid-parallel} workloads. This design mismatch leads to two critical failures. First, they lack effective multisource orchestration: trainer-colocated loaders prevent global coordination, while remote loading systems provide limited APIs for scheduling and mixing across heterogeneous data sources. Second, for these loaders, each parallelism-unaware dataloader instance is a heavyweight clone that must independently manage the data access for the entire set of M data sources and K parallel ranks (CP/PP), leading to severe memory overhead that scales with source diversity and parallelization size.

To address these limitations, we present \sys{}, a disaggregated data processing framework designed for large-scale foundation model 
workloads that require two core capabilities: global data orchestration and multisource scalability. 
The key insights behind \sys{} are the disaggregation of the multisource data preparation and the final delivery. It decomposes multisource preprocessing into specialized roles—dedicated Source Loaders for sample-level transformations (e.g., JPEG decoding) and aggregating Data Constructors for batch-level operations (e.g., tensor padding, splitting)—to eliminate redundant source-level and parallelism-related data access, also mitigating connectivity overhead. It introduces a programmable data plane with DGraph, a stateful dataflow graph tracking dependencies and lineage for each data source, and ClientPlaceTree, a hierarchical topology model enabling hybrid parallelism-aware scheduling, to support declarative cross-module multisource data strategy definition. It further employs source auto-partitioning and mixture-driven auto-scaling to dynamically optimize worker allocation for heterogeneous preprocessing demands.

Our core contributions are summarized below:

$\triangleright$ \textbf{Disaggregated multisource preprocessing architecture}: We design a distributed actor-model-based preprocessing pipeline that eliminates redundant data access and memory overhead in LFM multisource data preprocessing (Sec.~\ref{sec:sys_overview})

$\triangleright$ \textbf{Declaritive Load-time Data Orchestration}: The DGraph and ClientPlaceTree abstractions enable hybrid parallelism-aware data orchestration with minimal user coding effort (Sec.~\ref{sec:data_programming}).

$\triangleright$ \textbf{Adaptive Multisource Scaling}: We introduce scalable algorithms that dynamically optimize CPU utilization for data preprocessing 
based on heterogeneous source preprocessing costs and evolving data mixing ratios (Sec.~\ref{sec:auto_scale}).


$\triangleright$
On clusters of up to 4096 GPUs, \sys{} improves multimodal VLM training throughput by 4.5$\times$ and reduces CPU usage by 13.5$\times$ versus data parallel baselines. (Sec.~\ref{sec:eval}).

\section{Background and Motivation}\label{sec:motivation}

\begin{figure*}[t]
    \centering
    \includegraphics[width=\linewidth]{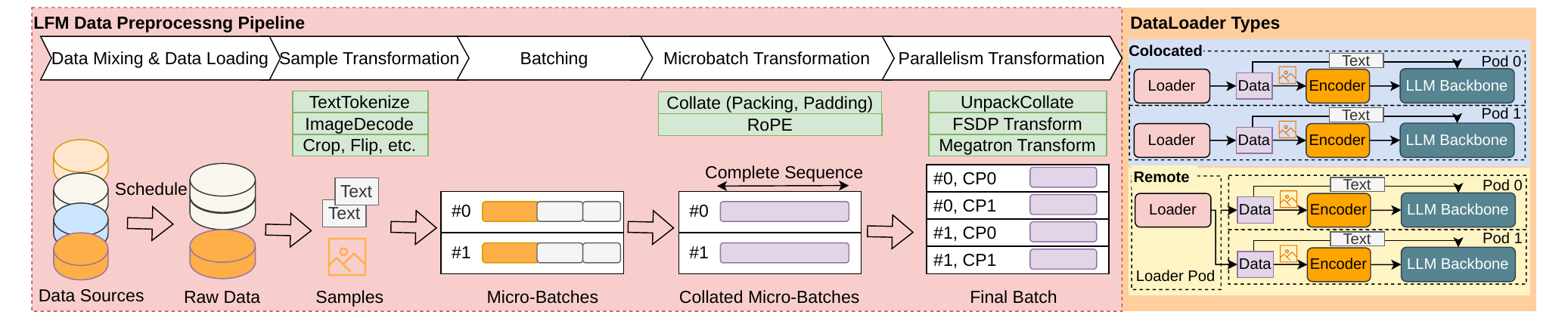}
    \vspace{-8mm}
    \caption{Left: Dataflow of Vision Language Model pretraining. Right: Two types of dataloaders.}
    \label{fig:dataflow}
    \vspace{-2mm}
\end{figure*}

We depict and analyze the production input data preprocessing pipeline for a Visual-Understanding-Language Model (VLM). Our analysis reveals the challenges and opportunities inherent in the data processing framework for LFM training.

\subsection{LFM Input Data Preprocessing}\label{sec:preprocess}
As shown in Fig.~\ref{fig:dataflow}, data from different datasets is mixed and loaded from cloud or distributed file systems, such as Amazon S3~\cite{amazon_s3} and HDFS~\cite{hdfs}. 
On-the-fly transformations are conducted to convert input training samples into tensors that are ready for training. 
This "last-mile" data conversion is performed by online data processing frameworks, commonly referred to as dataloaders, such as the PyTorch DataLoader~\cite{torch_dataloader}. 

\noindent 
\textbf{Data Mixing and Data Loading.}
In LFM training, samples from different data sources are combined in proportions to formulate a \textit{data mixture}~\cite{gemini}, resulting in a comprehensive, inclusive training corpus. 
The mixing ratio adjustment can be performed in a scheduled way, like with curriculum learning paradigms in reinforcement learning~\cite{soviany2022curriculumlearningsurvey}, warmup~\cite{li2022stabilityefficiencydilemmainvestigatingsequence} and staged training~\cite{gemini} techniques, or in a dynamic manner, where the sampling weights of the data sources vary in response to evolving training status (such as loss and entropy~\cite{ADO, skill-it}). The dataloader utilizes the sampling 
ratios of the mixing schedule to load data proportionally from different datasets (sources).


\noindent
\textbf{Sample Transformations}
are applied to each data sample, 
converting the data format and improving the data quality~\cite{tf_data_service, pecan, zhao2024cedar}. For example, text tokenization and image decoding convert raw text and image data into trainable representations (e.g., text tokens and normalized RGB tensors). Similarly, the \textit{crop} transformation standardizes image dimensions by cropping and resizing inputs to a fixed resolution.

\noindent
\textbf{Microbatch Transformation.}
After batching sampled data into microbatches, microbatch transformations collate samples to standardize the input shape. Packing merges fragmented subsequences into complete sequences with segmented masks~\cite{packing, navit}. Padding aligns variable-length sequences by adding dummy tokens~\cite{packing, navit}. Finally, positional embedding transformations like RoPE~\cite{rope} are applied to obtain complete context information.

\noindent
\textbf{Parallelism Transformation.}
To effectively parallelize LFM training with increasing data and model sizes, hybrid parallelism strategies have been adopted. 
Hybrid parallelisms significantly influence how 
model training consumes training data.
In data parallelism (DP)~\cite{dp_1, dp_2,dp_3}, microbatches are partitioned across training devices, and each device independently processes its local data using its model replica. Context parallelism (CP)~\cite{ringattention, megatron-3} partitions and scatters the input sequence; 
each device within a CP group consumes a part of the input sequence. 
Tensor parallelism (TP)~\cite{megatron_lm_train} performs intra-operator partitioning, with each device within a TP group receiving the same input.  
Pipeline parallelism (PP)~\cite{huang2019gpipe, narayanan2019pipedream} segments the model layers into stages, where only the first stage (PP0) 
obtains all microbatches, and intermediate results are exchanged between consecutive stages through peer-to-peer (P2P) communications. These replication and partitioning relationships between input data and hybrid parallelism schemes are encoded as parallelism transformations, applied after the microbatch transformation to ensure each client receives the correct input data.

Multimodal LFMs~\cite{DistMM, DistTrain} further complicate the input data processing pipeline with \textbf{heterogeneous colocation}. Multimodal LFMs are composed of more than one module (encoders, backbone, etc.), and each module may process different parts of the input data and employ different hybrid parallelism schemes. For example, 
a VLM can employ pure data parallelism for the vision encoder (e.g., ViT~\cite{vit}) training 
and 4D (PP-DP-CP-TP) parallelism for LLM backbone (e.g., LLaMA~\cite{llama3, meta2025llama4}) training. 
The parallelism difference for different data within a batch requires careful parallelism transformation programming.

\noindent
\subsection{Colocated and Remote Dataloader}
As shown on the right of Fig.~\ref{fig:dataflow}, conventional ML training frameworks (e.g., Torch~\cite{torch_dataloader}) typically colocate the dataloader with the training process. 
We observe that mainstream LFM training frameworks, such as Megatron-LM~\cite{megatron_lm_train}, DDP~\cite{DDP}, FSDP~\cite{torch_fsdp}, and MegaScale~\cite{jiang2024megascale}, still adhere to this practice.
One advantage of colocated dataloaders is that they share the same sharding configuration (e.g., data parallelism) and CPU resources with the training process, eliminating the need for additional configuration. However, their rigid loader setup limits their ability to right-sizing resources like CPU and DRAM for efficient data processing.  


\begin{figure*}[!ht]
    \centering
    \begin{subfigure}[t]{0.48\linewidth}
        \centering
        \includegraphics[width=\linewidth]{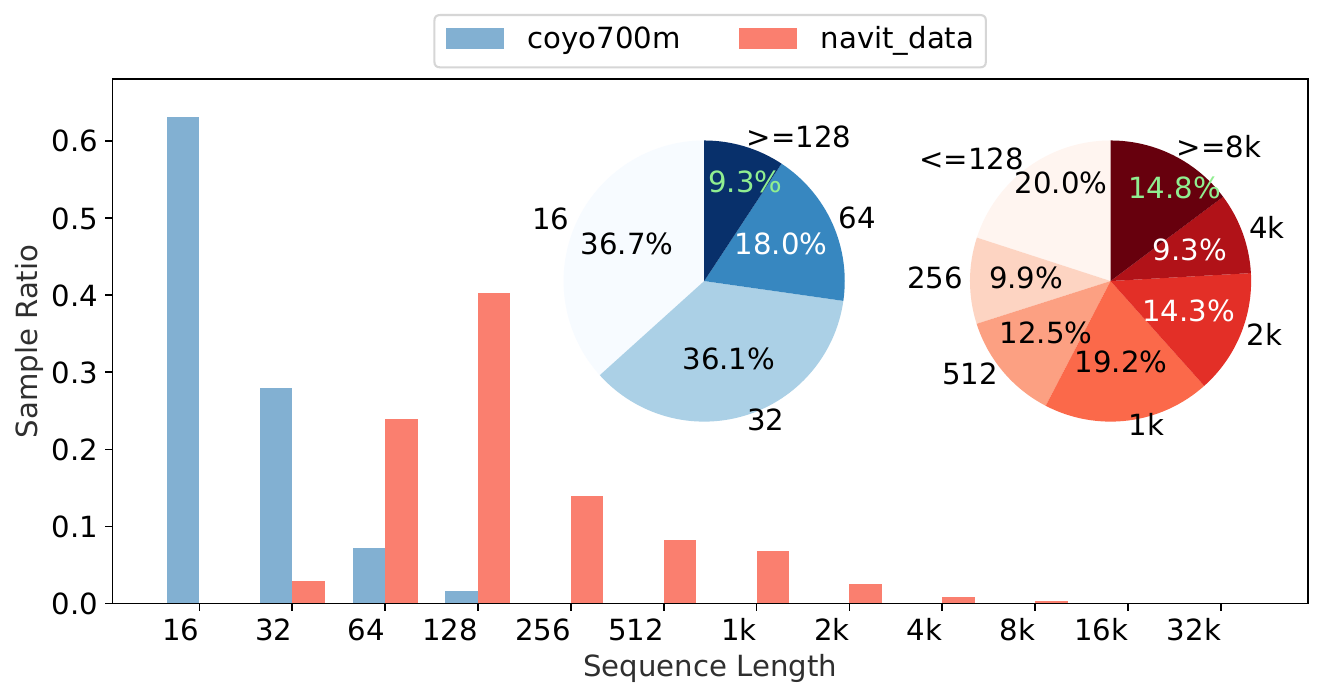}
        \caption{Text token distribution} 
    \end{subfigure}
    \hfill 
    \begin{subfigure}[t]{0.48\linewidth}
        \centering
        \includegraphics[width=\linewidth]{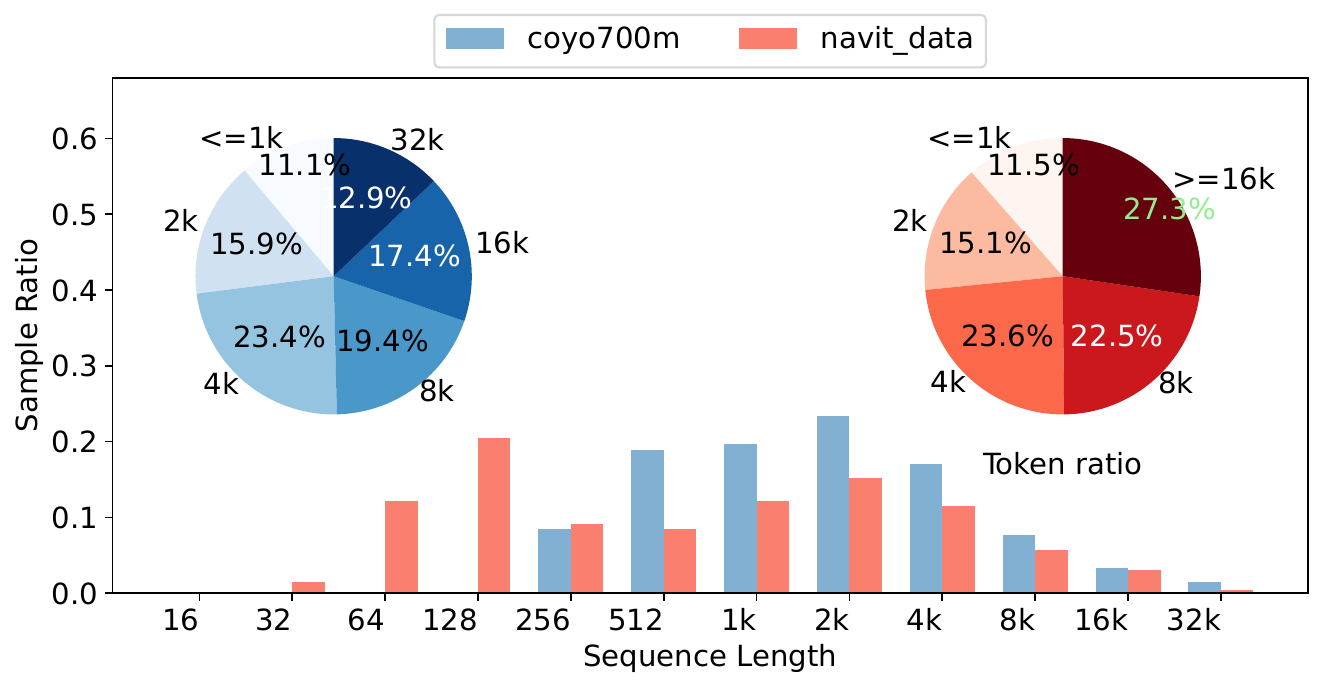}
        \caption{Image token distribution} 
    \end{subfigure}
    \caption{Skew token distribution in two datasets. Pie: total training token percentages by length; bar: sample ratios (e.g., short sequences dominate samples but long contribute more tokens). Left: Text token. Right: Image token.}
    \label{fig:sample-distribution}
\end{figure*}

Remote dataloaders, exemplified by Meta’s Data Processing Service (DPP)~\cite{DPP} and Google’s tf.data service~\cite{tf_data_service}, offload data preprocessing to disaggregated CPU workers over the accelerator nodes. 
This architecture enables elastic scaling of data preprocessing throughput. Recent studies—Cachew~\cite{graur2022cachew}, FastFlow~\cite{fastflow}, Cedar~\cite{zhao2024cedar}, and Pecan~\cite{pecan}—further optimize resource efficiency by orchestrating heterogeneous CPU resources across local and remote nodes. Nonetheless, all the existing dataloader systems fail to meet the new demands of LFM training, to be detailed in the following.

\subsection{LFM Data Preprocessing Requirements}\label{sec:requirements}

\begin{figure}[t]
    \newlength{\subfigheight}
    \setlength{\subfigheight}{5cm} 

    \centering
    \begin{subfigure}[b]{0.48\textwidth}
        \centering 
        \includegraphics[height=\subfigheight, keepaspectratio]{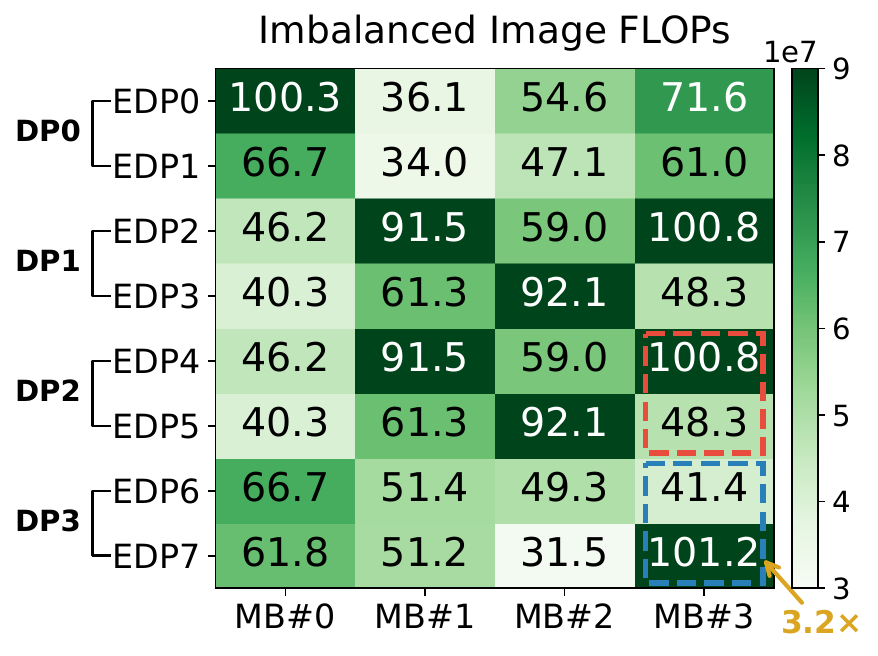}
        \caption{Image flops heatmap}
        \label{subfig:image_flops}
    \end{subfigure}
    \hfill 
    \begin{subfigure}[b]{0.48\textwidth}
        \centering 
        \includegraphics[height=\subfigheight, keepaspectratio]{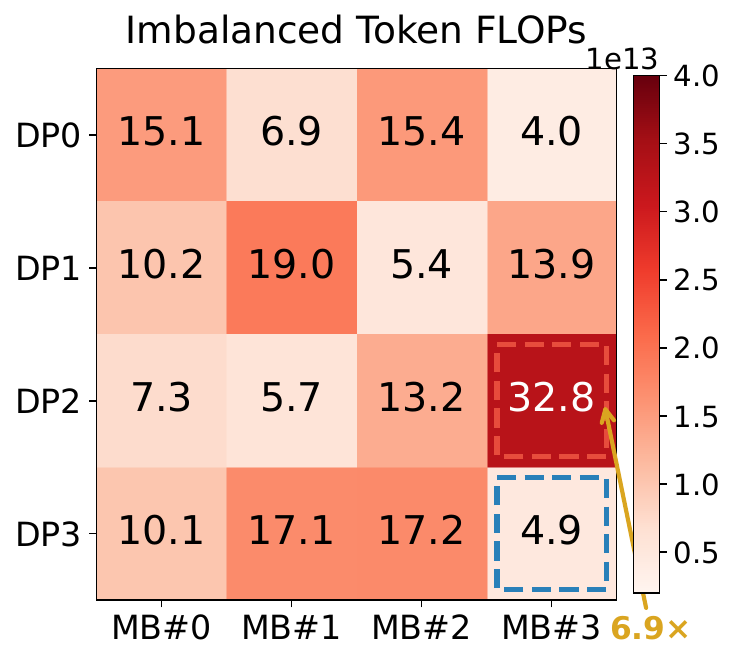}
        \caption{Token flops heatmap}
        \label{subfig:token_flops}
    \end{subfigure}

    \vspace{3mm} 
    \caption{Computational imbalance across microbatches. EDP$x$ denotes the encoder data parallel rank $x$, DP$x$ denotes the data parallel rank $x$, and MB\#$x$ represents the $x$-th microbatch.}
    \label{fig:imbalance_microbatch}
\end{figure}

\noindent
\textbf{Multisource Data Orchestration.} 
Data scheduling remains consistently necessary due to data heterogeneity under sophisticated training parallelism schemes.
VLMs construct training samples by interleaving encoded images and text tokens. Each input sample comprises an image-text pair: images are split into patches and encoded into tokens via a visual encoder, while text labels are tokenized separately. These token streams are then interleaved to form complete training sequences.
Fig.~\ref{fig:sample-distribution} shows the token distributions in the open-source \textit{coyo700M}~\cite{kakaobrain2022coyo-700m} dataset and our production \textit{navit\_data} dataset, where "text" represents text tokens, and "image" indicates the number of 16×16 (\textit{coyo}) and 14×14 (\textit{navit}) image patches~\cite{navit}.
 Both distributions are significantly skewed. In \textit{coyo700M}, 98.23\% of samples contain text sequences $\leq 64$ tokens, while the top 1.62\% of longer sequences ($>64$ tokens) account for 9.3\% of all tokens. Similar skewness occurs in image subsequences. Such skewness manifests two critical computational challenges:

\noindent \textit{(1) Intra-module imbalance.}
The quadratic time complexity of the attention operation ($O(l^2)$ with $l$ as sequence length)~\cite{dao2022flashattention} induces a significant computational disparity between microbatches containing subsequences of varying lengths. 
Fig.~\ref{fig:imbalance_microbatch} shows 
this disparity, where we benchmark an 8-card VLM training trial, colocating encoders and the backbone, employing Encoder Data Parallel (EDP) with size 8 to distribute image data across all GPUs, while the backbone utilizes hybrid parallelism with $DP = 4$ and $TP = 2$. The maximum microbatch FLOPs observed are $ 3.2\times$ and $6.9\times$ higher than the minimum for images and complete sequences, respectively, as indicated by the yellow arrows in Fig. \ref{fig:imbalance_microbatch}.


\noindent \textit{(2) Inter-module imbalance.}
Input data distributions can vary significantly across modules. As shown in Fig.~\ref{fig:imbalance_microbatch}, even for the same microbatch, there exists a substantial divergence in the token distributions between raw image patches and the final collated sequence. Given the heterogeneous parallelisms of collocated modules on the same GPU, Data Parallel rank $\#3$ (DP3) for the LLM backbone corresponds to Encoder Data Parallel ranks 6 and 7 (EDP6, EDP7). The computational complexity of image tokens across microbatch $\#3$ is balanced between every two consecutive DP ranks (140e7); however, as highlighted by the red and blue dashed outlines, DP2’s MB$\#3$ workload ($32.8 \times 10^{13}$ FLOPs) is significantly larger than DP3’s MB$\#3$ ($4.9\times 10^{13}$ FLOPs), indicating a severe imbalance. Moreover, when considering only the encoder data parallel ranks for MB$\#3$, the FLOPs across DP ranks remain unbalanced and necessitate adjustment. To address this, data scheduling strategies should be tailored distinctly for different modules to homogenize their workloads.  


Additionally, existing solutions often leverage all-to-all communication protocols at the model layer to aggregate information for load balancing input batches (e.g.,~\cite{DistTrain}). However, this approach introduces four fundamental challenges: increased activation size, elevated communication overhead, scalability bottlenecks, and tight coupling of load balancing logic with model-layer codebases, which risks contaminating the core computation workflow. We argue that load-time data orchestration presents a compelling alternative: an LFM data preprocessing system should proactively balance data from heterogeneous sources across modules before model ingestion.

\begin{figure}[t]
    \centering
    \includegraphics[width=\linewidth]{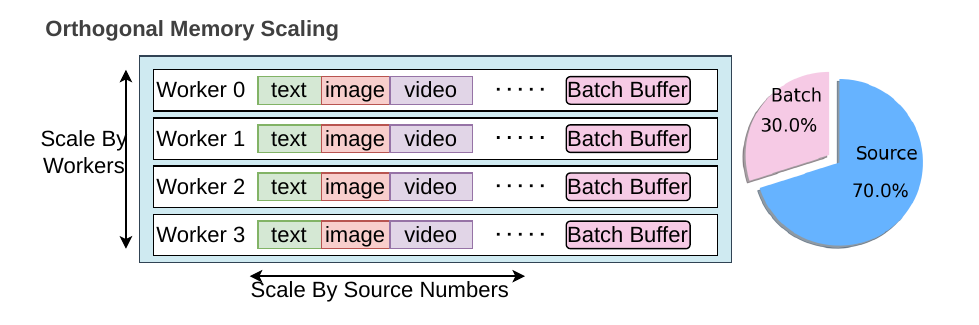}
    \vspace{-8mm}
    \caption{Orthogonal memory scaling by source and worker counts in multi-source preprocessing. Pie charts show that source-related memory dominates the usage. 
    }
    \label{fig:mem_scaling}
    \vspace{-4mm}
\end{figure}

\noindent
\textbf{Multisource Scalability.} 
The multisource nature of LFM training data exerts substantial memory pressure on the preprocessing pipeline. As demonstrated in Fig.~\ref{fig:mem_scaling}, our production-scale LFM training trials reveal that when the per-DP training batch size remains moderate, the memory footprint of replicated file access states for data sources constitutes over 70\% of the total memory consumption during data preprocessing. This multisource memory overhead arises from two orthogonal scaling dimensions.

\begin{figure}[t]
    \centering
    \begin{subfigure}[t]{0.48\linewidth}
        \centering
        \includegraphics[width=\linewidth]{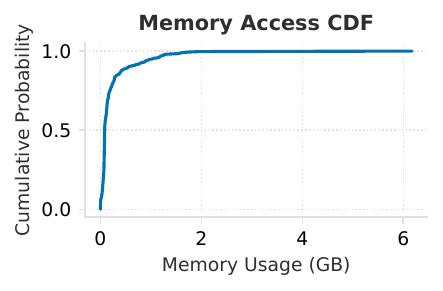}
        \caption{File access states memory}
    \end{subfigure}
    \hfill 
    \begin{subfigure}[t]{0.48\linewidth}
        \centering
        \includegraphics[width=\linewidth]{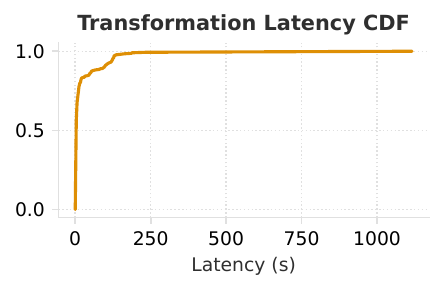}
        \caption{Transformation latency}
    \end{subfigure}
    
    \vspace{-4mm}
    \caption{Cumulative distribution function (CDF) of 100 samples across production source datasets.}
    \label{fig:source_dist}
    \vspace{-4mm}
\end{figure}

\noindent \textit{(1) Source Scaling.} Modern LFMs achieve task generality through source datasets aggregated from diverse domains~\cite{llama3, ds_llm, data_mix_laws}. In production systems, training jobs process hundreds to thousands of distinct source files that collectively define the global data mixture. Each file introduces a fixed per-source memory overhead. For instance, practitioners commonly store training data in columnar formats such as Parquet~\cite{parquet_apache}, which optimize compression ratios and access locality through feature grouping~\cite{peper}. Parquet files partition data into row groups (512MB–1GB storage units~\cite{parquet_apache}). During reads, a client first establishes a dedicated socket to the file, loads metadata (e.g., footers, schema information) into memory, and executes queries over row groups using buffers~\cite{columnar_format}. This design inherently leads to linear memory overhead scaling: independent file states (sockets, metadata, buffers) per source cause memory usage to grow proportionally with the source number. 

\noindent \textit{(2) Worker Scaling.} To prevent GPU idling, 
practitioners must optimally size dataloader worker processes to match data transformation bottlenecks. 
Each worker process maintains its execution context and prefetch buffer~\cite{torch_dataloader}, and thus the memory consumption scales with the number of workers. This issue is exacerbated by transformation time heterogeneity across modalities. For instance, text tokenization is lightweight, whereas image decoding (e.g., RGB conversion via PIL~\cite{PIL}) and video processing (e.g., keyframe extraction~\cite{keyframe}) are computationally intensive. Even for the data of the same modality, e.g., images, patches~\cite{navit} for variable-resolution images introduce order-of-magnitude cost differences. As a result, transformation latencies exhibit severe skewness across sources, as shown in Fig.~\ref{fig:source_dist}. Critically, when faced with multiple data sources, loader workers must dynamically resize to match the throughput of the slowest transformation pipeline. This prevents vibrating feeding rates that could otherwise introduce GPU idle time. However, this forces dataloaders to scale workers based on worst-case latency demands, necessitating over-provisioning for faster pipelines.  

These observations underscore the need for multisource-scalable preprocessing. An effective solution must address both (1) horizontal memory growth due to increasing sources and (2) vertical memory amplification from worker scaling under heterogeneous, time-varying transformation latency.

\subsection{Challenges For Preprocessing Requirements}\label{sec:challenge_preprocess}

\begin{figure}[t]
    \centering
    \includegraphics[width=1.02\linewidth]{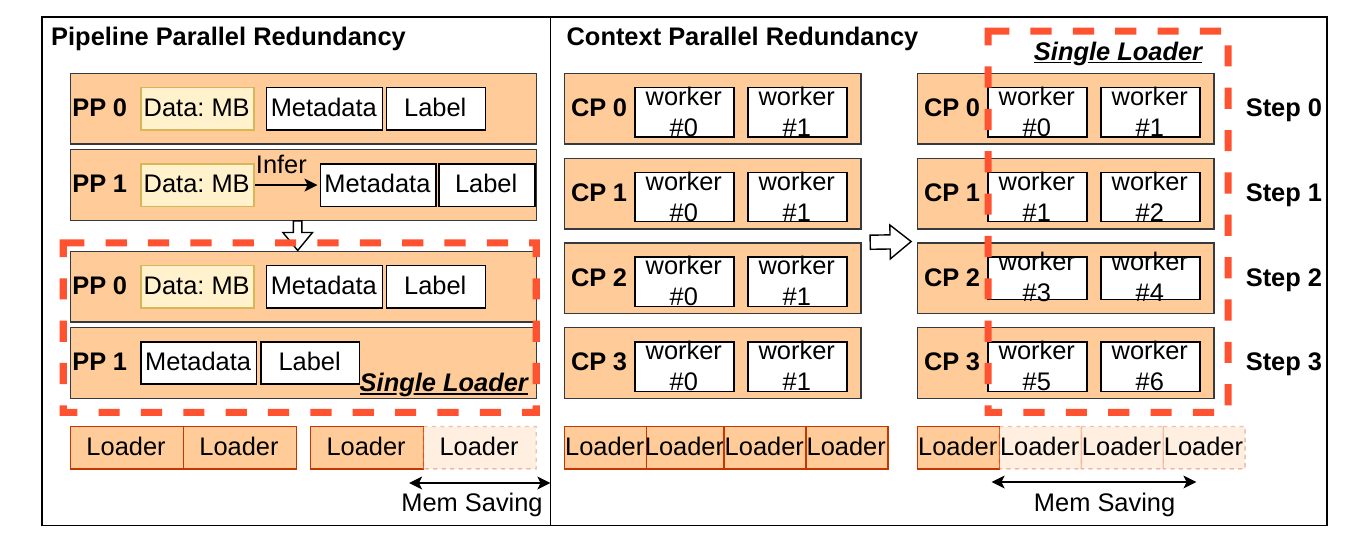}
    \vspace{-6mm}
    \caption{Optimizing Loader Parallelism Redundancy for Memory Saving. Left: share one loader instance result for per-stage loaders in the pipeline parallelism. Right: share one loader instance for per-rank loaders in context parallelism.}
    \label{fig:p-redundancy}
    \vspace{-5mm}
\end{figure}
In addition to the above requirements, 
an efficient design of an LFM multisource data preprocessing framework faces the following challenges. 

First, the dynamic data mixing of LFM training alters data source sampling ratios at runtime, necessitating the data orchestration panel to express and adapt to runtime data source sampling patterns. Additionally, it mandates that multisource scalability solutions not only support arbitrary mixing strategies but also automatically adjust the source mixing ratios as training progresses, given that per-source transformation latency evolves non-uniformly over time.

Next, as elaborated in Sec.~\ref{sec:preprocess}, ranks in context parallelism and pipeline parallelism require only partial input data for model execution. In the absence of coordination, each rank independently instantiates a full dataloader to load complete batches for data partitioning and metadata retrieval (Fig.~\ref{fig:p-redundancy}). This leads to redundant loader instances that scale with the number of CP/PP ranks, exacerbating memory overhead. To mitigate this, the framework must tightly integrate data orchestration with hybrid parallelism, eliminating redundant data access and memory occupation by sharing ephemeral, parallelism-aware preprocessed samples across ranks.

Finally, LFM training may dynamically adjust GPU allocations at runtime—elastic scaling (addition/deletion), redeployment, resharding, or failure events~\cite{wagenlander2024tenplex, wan2024bytecheckpointunifiedcheckpointingllm}—which requires persistent data services to maintain stable data feeding speeds. These scenarios demand built-in mechanisms in the data proprocessing framework for resilient connectivity and fault tolerance to ensure uninterrupted data delivery.

\subsection{
Opportunities}
Despite their advances, existing data loading systems are ill-equipped for the unique demands of large-scale, multi-source LFM training. The core optimizations offered by state-of-the-art remote dataloaders~\cite{pecan, graur2022cachew} are fundamentally misaligned with multi-source LFM workloads. Their focus on caching offers little benefit in typical single-epoch training scenarios, and their emphasis on offloading CPU-bound transformations is less critical given the long training iteration time and abundant host CPU cores.

Consequently, these systems fail to address the primary bottlenecks in this domain: 1) the lack of expressive APIs to orchestrate the complex, dynamic data mixing required by hybrid parallelism, which forces developers into manual, error-prone implementations; and 2) severe memory redundancy that scales linearly with the number of concurrent data sources and loader instances, resulting in unsustainable resource costs. Furthermore, most systems lack robust fault tolerance, making large-scale training runs fragile. 

\sys{} addresses these issues with three architectural innovations. It employs an actor-model architecture to split preprocessing into Source Loaders and Data Constructors, improving scalability and eliminating redundant access in hybrid parallelism and multisource setups. A centralized data plane with declarative interfaces simplifies cross-module data scheduling. Multilevel source auto-partitioning and mixture-driven scaling optimize resource use.  \sys{} also ensures uninterrupted data service by incorporating shadow dataloaders for high-availability fault tolerance and an elastic resharding mechanism to dynamically adapt to training parallelism changes. A comprehensive
discussion with related works is provided in Appendix~\ref{app:related-work}.


\section{\sys{} Overview}~\label{sec:sys_overview}
\begin{figure}
\centering
\includegraphics[width=0.65\linewidth]{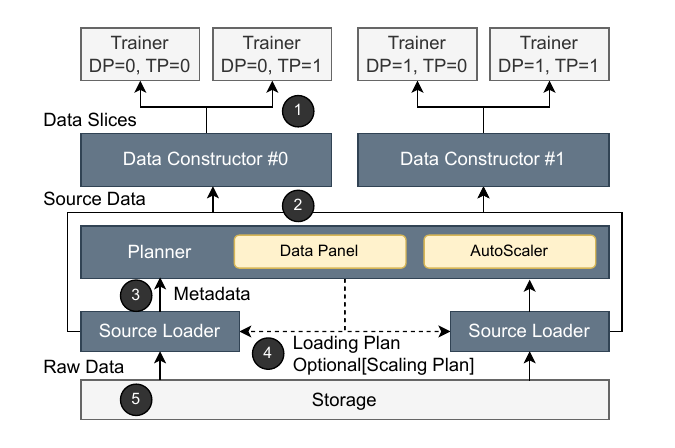}
\caption{Architecture of \sys{}. Solid lines denote data paths; dashed lines represent control paths.}
\label{fig:sys_design_arch}
\end{figure}

\sys{} is a data preprocessing framework designed for unified orchestration and resource-efficient delivery of multisource training data. As shown in Fig.~\ref{fig:sys_design_arch}, the framework employs a hierarchical architecture with three core components, explicitly designed to address the scalability and orchestration challenges:

\noindent
\textit{Source Loader (Addressing Source Redundancy \& Source Heterogeneity).} To resolve memory constraints caused by source scaling, Source Loaders function as dedicated actors for specific data sources. By continuously ingesting data and applying sample-level transformations within isolated processes, they decouple file access states from worker processes, effectively eliminating the redundancy of file handles and metadata across the system.

\noindent
\textit{Data Constructor (Addressing Parallelism Redundancy).} 
Acting as the data sink for ranks (e.g., Data Parallel group), the Data Constructor aggregates outputs from Source Loaders to perform batch-level and parallelism transformations. This design enables seamless data sharing: ranks in the same CP group can generate partitions from identical batches, while PP stages can selectively exclude unnecessary metadata, significantly reducing memory and I/O overhead. The Larger cluster with inter-node parallelism requires finer-granular Data Constructor and selective broadcasting (See Sec.~\ref{sec:ft_sys}) to reduce synchronization cost.

\noindent 
\textit{Planner (Addressing Unified Orchestration \& Scaling).} The Planner orchestrates system-wide behavior through two key functions: (1) {Plan Generation}, which synthesizes loading plans by aggregating lightweight metadata (e.g., sample indices, sequence lengths) from Source Loaders and applying user-defined orchestration strategies; and (2) {Auto-Scaling}, which triggers resource resizing (Sec.~\ref{sec:auto_scale}) to adapt to preprocessing workload fluctuations and cost-efficiency targets.


\noindent
\textbf{Deployment.} \sys{}'s components are deployed entirely in CPU memory. At bootstrap, the user defines a strategy using the \sys{} programming model (Sec.~\ref{sec:data_programming}), which the Planner's Data Panel then instantiates. This strategy provides policy-driven control over dataflows. Based on the trainer device topology, \sys{} provisions Data Constructors and trainer clients. It also automatically partitions source datasets into multiple Source Loaders during instantiation, using journalized profiling results. To adapt to dynamic workload changes, the AutoScaler adjusts resource allocation in response to fluctuating mixing ratios (Sec.~\ref{sec:auto_scale}).

\noindent
\textbf{Workflow}.
The runtime data preprocessing follows a pull execution model~\cite{valcano, flink, kafka}. Each component (Data Constructor, Source Loader, Planner) maintains its own task queue. The workflow proceeds as:
\circlednumber{1} A trainer-side client requests data from its Data Constructor.
\circlednumber{2} The Data Constructor triggers positional data fetches from all Source Loaders.
\circlednumber{3} Source Loaders consult the Planner to generate new loading plans.
\circlednumber{4} The Planner synthesizes new plans by collecting buffer metadata (e.g., sample indices, source signatures, sequence length) from Source Loaders. Finalized plans direct Source Loaders to prepare samples, pop them from the read buffers, stage them in queues, and signal a resource scaling plan if necessary. \circlednumber{5} Source Loader reads new samples from the distributed storage and populates them into the buffer.


\section{Data Orchestration Panel}\label{sec:data_programming}
\begin{figure*}[!t]
    \centering
    \includegraphics[width=\linewidth]{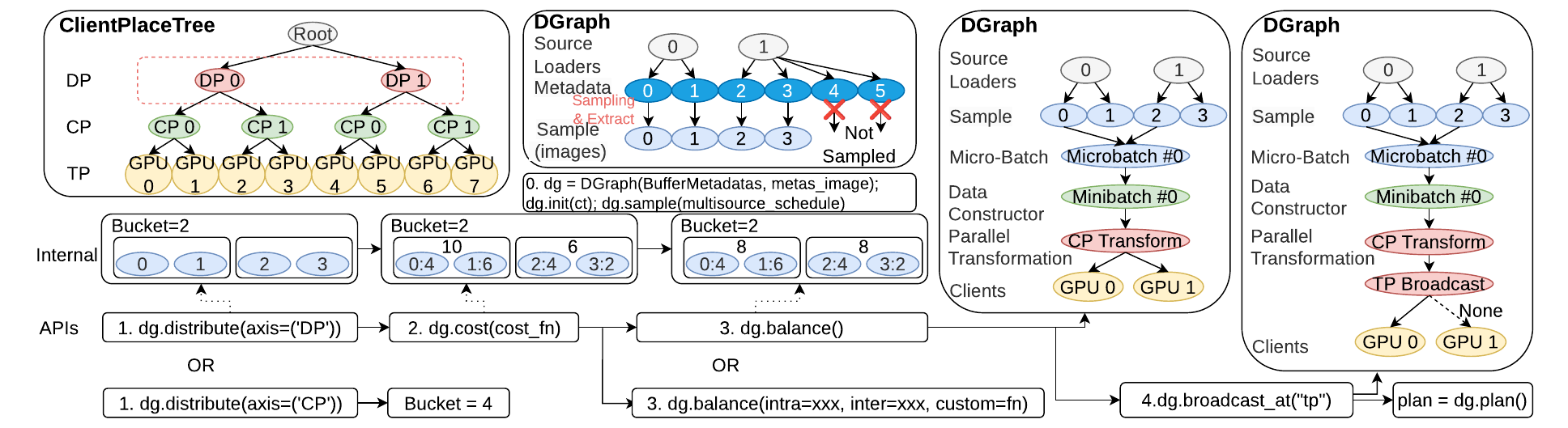}
    \caption{\textbf{Multi-source data balancing under hybrid parallelism (DP=2, CP=2, TP=2).} 
    \textbf{(Top)} \textit{DGraph} tracks sample lifecycles while \textit{ClientPlaceTree} defines the logical device mesh. 
    \textbf{(Middle)} \textit{Internal} state: ellipses represent samples (e.g., ``0:4'' denotes ID 0 with cost 4); rounded rectangles are microbatch bins with aggregated costs shown above. 
    \textbf{(Bottom)} API workflow: \texttt{distribute} creates $n$ buckets based on the tree level (e.g., $n=4$ for \texttt{axis='CP'}); \texttt{cost} maps metadata to overhead; \texttt{balance} redistributes samples into bins. Data (4, 5) are excluded based on the sampling results.}
    \label{fig:programming_sample}
\end{figure*}
\subsection{Abstraction}
Existing dataloader frameworks like tf.data.service~\cite{tf_data_service} provide distributed execution APIs but lack native abstractions for two critical requirements: (1) multisource representation for data within the same batch, and (2) trainer-side hybrid parallelism scheme integration (e.g., 4D parallelism, PP, DP, CP, and TP). These gaps complicate unified implementation of multisource and cross-module data orchestration.

To address these limitations, \sys{} introduces two core abstractions: \textit{DGraph} for source-aware data transition state tracking and \textit{ClientPlaceTree} for parallelism strategy resolution, as 
exemplified in Fig.~\ref{fig:programming_sample}. 

\noindent
\textbf{\textit{DGraph}} is a state-tracking dataflow graph that models the lifecycle of training samples through explicit producer-consumer relationships. It is initialized by binding data samples to their respective Source Loaders. Upon plan generation, it associates each sample to its target Data Constructor.
Each node in \textit{DGraph} represents a training sample in a specific processing state, with directed acyclic edges encoding either data transformations or logical dependencies (e.g., microbatch grouping). Edges may be null if no state mutation occurs. \textit{DGraph} operates on lightweight metadata and delivers two core advantages: 
\textit{Unified multisource representation}, that allows creation of multiple source-specific graphs (e.g., test-image pair or text data from a specific source) from the same shared data dictionary through selective metadata specification; \textit{Orchestration transparency}, which visualizes dataflow states, dependencies, and transformations via directed acyclic edges in an interpretable way.

\noindent
\textbf{\textit{ClientPlaceTree}} is a logical representation of the trainer device mesh. It provides a clear view of how data is accessed by trainers at different ranks and in different communication groups while abstracting details such as device memory capacities from users. Presenting the topology as a tree allows for easy modifications to the structure and automates the generation of parallelism transformation. The cost of rebuilding the ClientPlaceTree is negligible; any changes to the device mesh parallelism~\cite{wagenlander2024tenplex} are dynamically reflected in the loading plan computation, as detailed in Sec.~\ref{sec:ft}. While typically inferred from the training configuration, the ClientPlaceTree allows users to override the default construction logic to implement custom behaviors, such as selective broadcasting in Sec.~\ref{sec:ft_sys}.


\subsection{Primitives}

As illustrated in Fig.~\ref{fig:programming_sample}, \sys{} initializes its data orchestration primitives using buffer metadata collected from Source Loader buffers and a \textit{ClientPlaceTree} encoding GPU allocations. Orchestration begins by specifying the modality metadata to create \textit{DGraph}, enabling \sys{} to encode either single or multiple modalities for load balancing. This modular approach is particularly valuable for models like VLMs, where multiple modules within the same process handle different data modalities at runtime, allowing separate specifications of balancing and transformation strategies. To facilitate this, \sys{} adopts a declarative interface. We define several primitives for \textit{DGraph} that allow users to describe the desired data distribution and transformation logic without managing low-level execution steps. These primitives can express most of our existing data orchestration strategies.

\noindent
\textbf{\texttt{\small mix(schedule)}} enables real-time source mixing through scheduled sampling. Users define a multisource schedule that generates the sampling weight of source datasets for each training step (at epoch, step, or substep granularity \cite{PPO}), determining the probabilistic selection of source data batches. Only sampled data participates in subsequent orchestration.

\noindent
\textbf{\texttt{\small distribute(axis, group\_size)}} specifies the axis along which data distribution occurs. This enables straightforward implementation of diverse partitioning strategies: (1) \texttt{axis=`DP'} partitions data into minibatches across data-parallel groups; \\(2) \texttt{axis=`CP'} treats $DP \times CP$ GPUs as uniform consumers for hybrid data parallelism~\cite{ge2025bytescaleefficientscalingllm}; (3) \texttt{axis=`WORLD'} distributes data across all ranks for the encoder module in VLMs with world-wide data parallelism. Once the distribution axis is selected, \sys{} automatically determines the parallelism transformations (e.g., CP transformations). The primitive creates $n$ buckets corresponding to nodes at the specified axis level in the ClientPlaceTree hierarchy. When \texttt{group\_size} is provided, the effective bucket count scales to $\ceil{\frac{n}{\texttt{group\_size}}}$, and samples are balanced within subgroups instead, reducing coordination overhead in super large clusters.

\noindent
\textbf{\texttt{\small cost(costfn)}} registers a cost function that estimates compute and memory overhead from sample metadata. Costs are automatically propagated to subsequent \texttt{\small balance} operations. Specifically, we model the encoder's cost as a function of the image sequence length, the dimensions of the embedding and MLP layers, and the model's depth. The cost for the language backbone is likewise modeled as a function of the total sequence length and key architectural parameters, such as the number of experts per token, vocabulary size, and hidden layer dimensions. We validate the latency fidelity of this cost model in Sec.~\ref{sec:ablation}.


 \noindent
\textbf{\texttt{\small balance(method, *)}} balances samples based on their computed costs. It further divides buckets into $m$ bins, where $m$ is the number of microbatches, and then applies the specified balancing method to distribute samples among these bins. We provide two candidate balancing methods: \texttt{greedybinpacking} and \texttt{karmarkar-karp}~\cite{karmarkar_karp}.  These can be applied at both intra-microbatch (bin) and inter-microbatch (bucket) levels~\cite{DistTrain}, or interleaving. To keep the global batch unchanged, users may optionally disable intra-microbatch reordering via configuration. 
User-defined balancing strategies (e.g., Zig-Zag, V-Shape, and other interleaved patterns) can be implemented via the framework's extension APIs.

\noindent
\textbf{\texttt{\small broadcast\_at(target\_dim)}} indicates to \textit{DGraph} that a broadcast operation exists on the trainer side along the specified dimension. For example, when TP0 broadcasts data to all TP ranks, this directive informs the Data Constructor to exclude $\rm{tp}>0$ clients from data fetching. This mechanism optimizes communication by preventing redundant data access.

\noindent
\textbf{\texttt{\small plan()}} dynamically generates optimized data-loading plans. Based on the plans, the Source Loader operates according to the plan to execute data mixing and scheduling, and then forwards prepared samples to the Data Constructor. The Data Constructor handles microbatch assembly and performs microbatch and parallelism transformations.

In addition to these primitives, we offer low-level programming interfaces with better flexibility, including \texttt{plan\_raw}, \texttt{loader\_do\_plan}, \texttt{constructor\_do\_plan} and \texttt{summary\_buffer}, which enable the user to program Planner, Source Loaders, and Data Constructor data consuming policies directly.

\subsection{Use Cases}\label{sec:use_case}
\begin{figure}[t]
    \centering
    \includegraphics[width=0.65\linewidth]{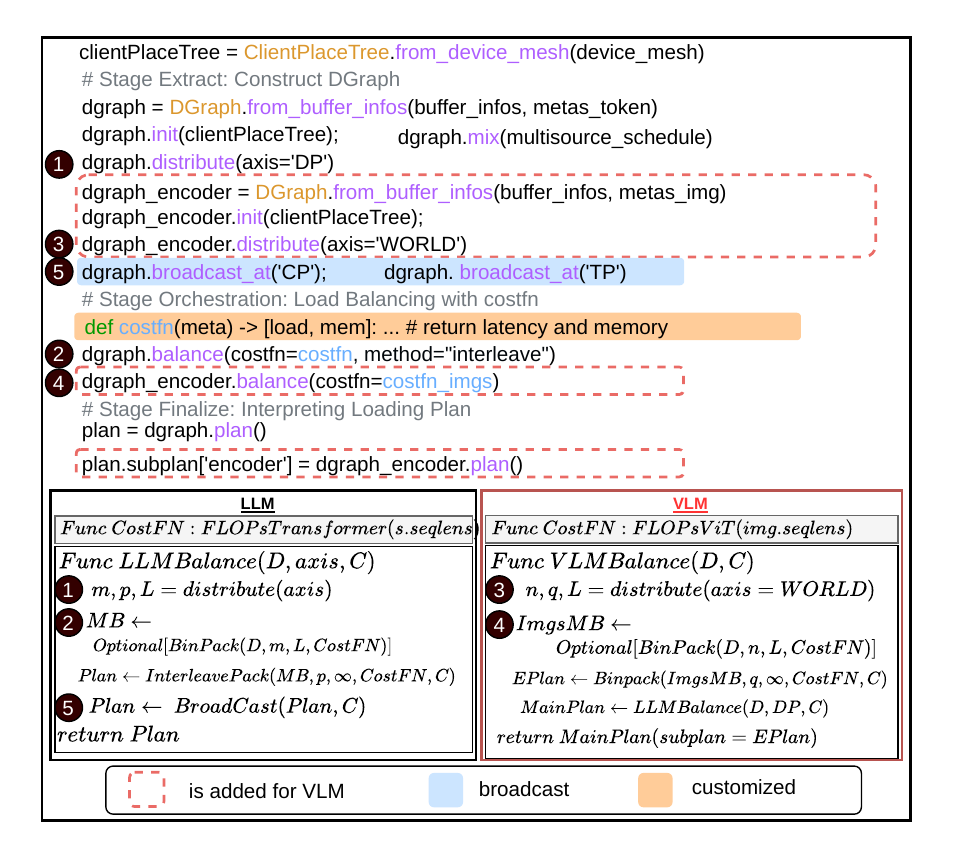}
    \caption{Strategy implementation of unimodal long-short sequence and multimodal vision language model pretraining task. Users can adapt to new training tasks by adding new lines and customizing limited functions.}
    \label{fig:prog_implementation}
\end{figure}

As shown in Fig.~\ref{fig:prog_implementation}, our APIs simplify the development of data orchestration strategies through a three-stage workflow: Extract, Orchestrate, and Finalize. To showcase the broad applicability and adaptability of our programming model, we analyze two representative use cases.
In the scenario of unimodal long-short sequence 4D-parallel training across data-parallel ranks, taking data $D$, and ClientPlaceTree $C$ as inputs, our framework achieves intra-module balancing (LLMBalance) using only seven lines of code, incorporating a cost model ($CostFN$). \circlednumber{1} First, it distributes data along the DP axis to determine the number of data-parallel partitions, per-DP microbatches, and memory limit $L$. \circlednumber{2} Next, it generates a resource consumption plan through balancing with greedy binpack. \circlednumber{5} If broadcast operations are included, the consumers are tailored.
For the VLM setup, an inter-module balancing strategy is implemented with just five additional lines of code. The image DGraph is inferred using the same buffer but different metadata.  We first distribute \circlednumber{3} and balance \circlednumber{4} images with data parallelism, and then combine with $LLMBalance$ to perform global balancing across the modules. 
These concise implementations 
demonstrate the expressiveness and versatility of our framework in diverse LFM training environments.


\begin{figure*}[t]
    \centering
    \includegraphics[width=\linewidth]{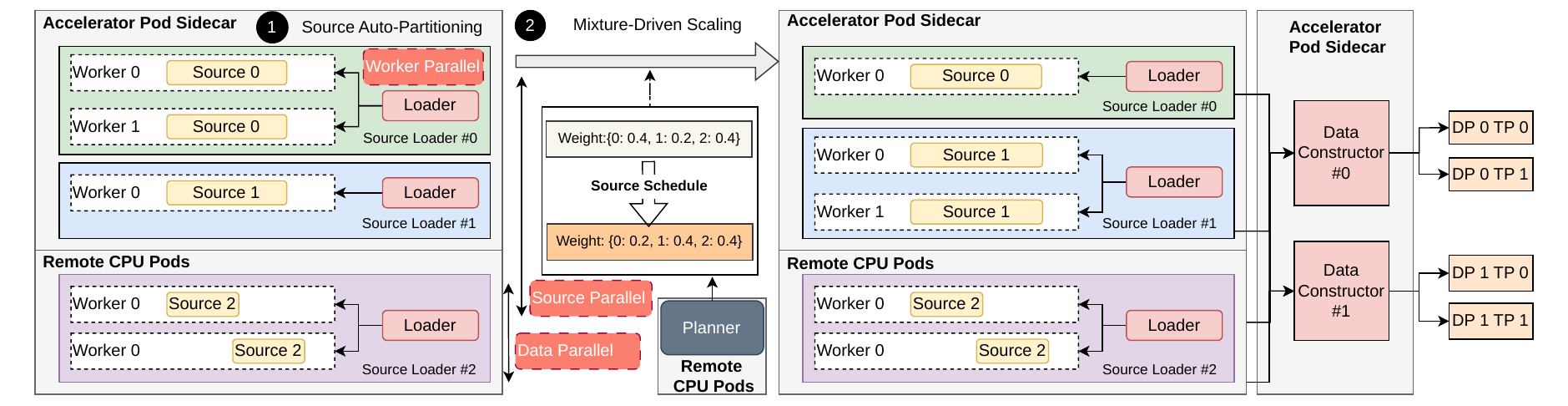}   
    \caption{Two-Phase AutoScaling. (1) Offline Source Auto-Partitioning, (2) Online Mixture-Driven Scaling. The number of workers for each source loader is adjusted according to the transformation cost of the corresponding dataset.}
    \label{fig:loader_role}
\end{figure*}

\section{MultiSource AutoScaler}\label{sec:auto_scale}

The AutoScaler in \sys{} addresses two core multisource scalability challenges: (1) partitioning and balancing heterogeneous preprocessing workloads and memory footprints across data sources, and (2) scaling of dynamic source mixing ratios. As shown in Fig.~\ref{fig:loader_role}, \sys{} employs a two-phase approach: \textbf{\circlednumber{1} Offline Source Auto-Partitioning:}
Sources are partitioned with loader parallel schemes to obtain Source Loader configurations. \textbf{\circlednumber{2} Online Mixture-Driven Scaling:}
At runtime, \sys{}'s AutoScaler in Planner dynamically scales and reshards Source Loaders in response to mixing ratio shifts, keeping low data preprocessing cost and resource efficiency under fluctuating demand.

\subsection{Source Auto-Partitioning}
We first analyze the parallelism schemes for partitioning the Source Loader, then introduce our scalable solution and elaborate its underlying rationale.

\noindent
\textbf{Dataloader Parallelism Schemes}. The end-to-end preprocessing cost is modeled as a tuple $(P, T, M)$, where $P$ denotes transformation latency, $T$ represents orchestration and data transfer overhead, and $M$ signifies memory footprint, comprising batch buffer ($M_b$) and per-source file access states ($M_d$). As shown in Fig.~\ref{fig:loader_role}, we analyze three parallelism strategies to distribute costs across loaders: (1) \textit{Worker Parallel} amortizes $P$ by staggering execution. Parallel workers in a loader initiate transformation $s-1$ steps ahead of their successors, where $w$ is the total worker count, enabling concurrent execution. (2) \textit{Source Parallel} reduces $M_d$
by partitioning data sources across multiple Source Loaders. Each loader worker maintains fewer file access states, decreasing memory pressure from metadata and I/O channels. (3) \textit{Data Parallel} reduces $P$ and $M_b$ by sharding training data across loaders~\cite{tf_data_service, pecan}, whose degree is bounded by the batch size. 

\noindent
\textbf{MultiLevel Source Partition}\label{s:autopartition}
Given heterogeneous transformation costs $\{P_1,..., P_k\}$ and memory footprints $\{M_1,..., M_k\}$ across $k$ data sources, we partition each source by default into multiple data-parallel actors, each containing worker-parallel workers with varying counts. The partition algorithm proceeds in three stages. (1) Source Clustering: We first sort all data sources in descending order of transformation cost $P_k$ and cluster them into $G$ source clusters. 
(2) Resource Level Construction: Using the ratio of mean transformation costs between the smallest and largest clusters, we estimate the number of workers for each source within a cluster. Available resources are calculated by subtracting resource allocations for the Data Constructor (estimated via fixed batch size) and Planner from total system resources. These available resources are divided by the total number of workers to form worker resource blocks. To prevent invalid data parallelism for global batches and physical pod resource overcommitment, we set upper bounds $w_{\text{src}}$ (per-source worker limit) and $w_{\text{actor}}$ (per-actor worker limit), and then generate actor counts (loader data parallelism degree) 
and worker counts for each resource level.
(3) Configuration Generation: We derive resource configurations for each source within clusters. When memory resources are insufficient for $M_k$, we adjust the number of source actors to satisfy memory constraints.

\subsection{Mixture-Driven Scaling} A key architectural advantage of \sys{} is the Planner’s centralized control over data mixture sampling, providing global visibility into cross-source mixing ratios (Fig.~\ref{fig:sys_design_arch}). This enables predictive autoscaling: as sampling weights evolve, the Planner dynamically adjusts Source Loader resources based on projected demand. When a source’s moving-average sampling weight exceeds a threshold for consecutive intervals, the Planner triggers the AutoScaler to: (1) create new Source Loader actors, (2) reshard data partitions live, and (3) integrate the scaled actors into the plan generation. Idle resources are reclaimed 
under declining demand.


\subsection{Design Rationale} 
We default to assigning one source per Source Loader actor, trading moderate communication for resource overhead, for two reasons. First, arbitrary mixing schedules complicate source grouping: maintaining stochastic consistency is challenging when sources in the same group exhibit divergent weight trajectories. Second, dedicated per-source actors eliminate most scaling needs from mixture changes, stabilizing feeding rates. While grouping remains suitable for static schedules, users can enable it in step (1) of source partition.  

Our focus on horizontal scaling via actor/worker counts utilizes underutilized CPU resources co-located with accelerators. Empirical studies \cite{hu2024characterizationlargelanguagemodel} and our trials show 75\% idle auxiliary CPU capacity under static allocations; actor-level management enables finer-grained control.

\section{Large-Scale Deployment}\label{sec:ft_sys}
\sys{} is implemented atop Ray~\cite{ray}, with 8763 lines of Python code. We discuss practical issues and solutions when deploying \sys{} at scale.


\begin{figure}[t]
    \centering
    \includegraphics[width=0.65\linewidth]{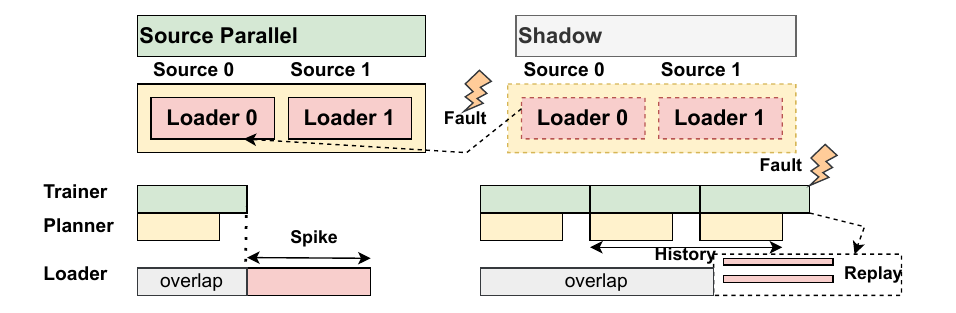}
    \caption{Failure Recovery: Shadow Loader, Replay with Differential Checkpointing Intervals.}
    \label{fig:shadow_loader}
\end{figure}

\subsection{Operational Adaptability}\label{sec:ft}
\sys{} incorporates mechanisms for fault tolerance and elastic resharding to ensure continuous data service.

\noindent\textbf{Fault Tolerance.} We decouple recovery mechanisms by component role.
Core coordinators (Planner, Data Constructor) leverage the Global Control Store (GCS) for state management and automatic restarts.
\sys{} uses persistent checkpointing and prefetch buffers to effectively mask this recovery latency.
For Source Loaders, the Planner detects failures via RPC timeouts or payload integrity checks (e.g., partial yields lacking end-of-stream signals).
As shown in Fig.~\ref{fig:shadow_loader}, upon detection, it promotes hot-standby ``Shadow Loaders'' for instantaneous failover.
To mitigate the latency of snapshotting large data buffers, we employ \textit{differential checkpointing}: Source Loaders snapshot at a lower frequency than the Planner, utilizing deterministic replay of the Planner state to bridge the gap upon failure.

\noindent\textbf{Elastic Resharding.} To support dynamic changes in training parallelism, \sys{} listens for notifications from the training framework. Upon receiving one, it immediately recalculates its data distribution plan for later incoming metadata and performs a fast resharding of resident data in Data Constructor to align with the new device topology.

\subsection{Deployment}
We employ the following deployment tricks.
\textbf{(1) Hybrid Deployment.} We adopt a hybrid model to balance efficiency and scalability (see Fig.~\ref{fig:loader_role}). The {Planner} runs on remote CPU pods for centralized scheduling. In contrast, {Source Loaders} and {Data Constructors} are primarily deployed as \textit{sidecars} within accelerator pods to utilize idle local CPU/memory resources~\cite{k8s_sidecar_2024}. They only scale out to remote CPU pods when sidecar resources become insufficient.
\textbf{(2) Transformation Reordering.} Inspired by Pecan~\cite{pecan}, we optionally defer image decoding to the Data Constructor, reducing communication data size.
\textbf{(3) Selective Broadcasting.} To tackle the high synchronization overhead of trainer-side client barriers in large clusters, we propose bottom-up selective broadcasting over ClientPlaceTree (e.g., broadcasting tensors within sub-communication groups such as TP and CP). This increases memory and communication costs but reduces the number of synchronized clients.

\begin{figure*}[!t]
\centering
\includegraphics[width=\linewidth]{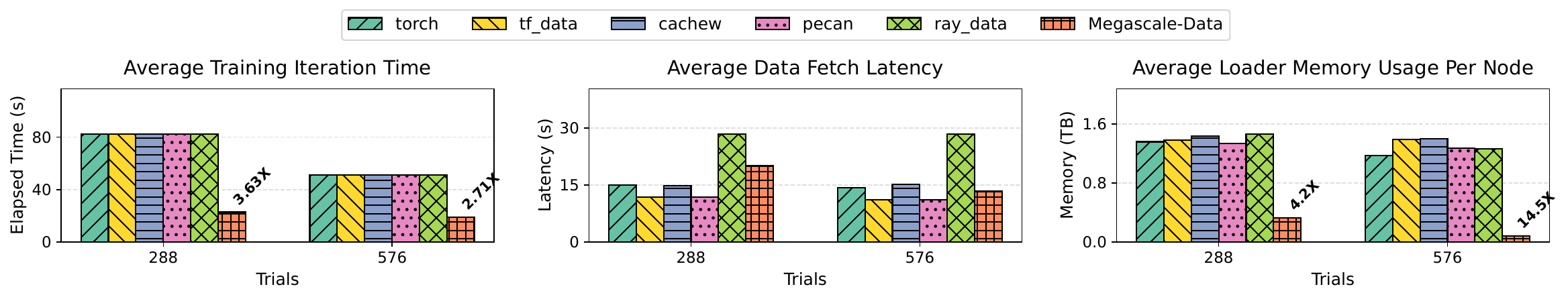}
\vspace{-8mm}
\caption{Comparison for data processing systems under different configurations. The parallelism strategies are ($TP=4, PP=8, DP=9$) for the 288-GPU trial and ($TP=4, PP=4, CP=4, DP=9$) for the 576-GPU trial. For all cases, we make data fetch latency fully overlapped with training time, i.e., not input-bound.}
\label{fig:main-loader}
\end{figure*}

\begin{table}[t]
    \centering
    \caption{Model configurations.}
    \resizebox{0.65\columnwidth}{!}{
        \begin{tabular}{llccc}
            \toprule
            & Model & \#Layers & \#Heads & Hidden Size \\
            \midrule
            \multirow{2}{*}{Encoder} 
            & ViT - 1B & 39 & 16 & 1408 \\
            & ViT - 2B & 48 & 16 & 1664 \\
            \midrule
            \multirow{3}{*}{LLM} 
            & Llama - 12B & 45 & 36 & 4608 \\
            & tMoE - 25B & 42 & 16 & 2048 (topk = 2) \\
            & Mixtral - 8×7B & 32 & 32 & 4096 (topk = 2) \\
            \bottomrule
        \end{tabular}
    }
    \label{tb:models}
\end{table}

\section{Evaluation}\label{sec:eval}
We evaluate \sys{} by quantifying its improvements in orchestration performance and multisource loading efficiency under large-scale deployment.

\subsection{Experiment Setup}\label{sec:exp_setup}

\noindent
\textbf{Models.} We evaluate \sys{} on Visual-Language Models (VLMs) using both dense and sparse Large Language Models (LLMs) as backbone, in combination with the Vision Transformer (ViT)~\cite{vit} as the image encoder. The details of the models are given in Table~\ref{tb:models}. For the dense model, we select the Llama3 series~\cite{llama3}. For sparse models (Mixture-of-Experts - MoE), we choose the Mixtral series~\cite{jiang2024mixtralexperts}. We also evaluate one production MoE model named tMoE.

\noindent
\textbf{Workloads.} 
We use two dataset groups (Fig.~\ref{fig:sample-distribution} in Sec.~\ref{sec:requirements}), namely \textit{coyo700m} and \textit{navit\_data}, consisting of 5 and 306 sources, respectively. We test \sys{} across various batch sizes, context lengths, and GPU numbers. We further scaled up to 4096 GPUs to benchmark scalability in Sec.~\ref{sec:ablation}.


\noindent
\textbf{Baselines.}
We evaluate \sys{} against state-of-the-art frameworks representing three data processing strategies: local processing (PyTorch DataLoader~\cite{torch_dataloader}, tf.data~\cite{tf.data}), remote processing (cachew~\cite{graur2022cachew}, Ray Data~\cite{ray_data}), and hybrid local-remote processing (Pecan~\cite{pecan}). For a fair comparison, we enable worker number auto-tuning for all data loaders to overlap data preprocessing time and model training iteration time with minimum CPU resources. By default, we enable the TP broadcasting for all data loaders. 

For orchestration evaluation, we gauge the orchestration efficiency of our solution in three scenarios: (1) Vanilla, a baseline system without any data scheduling; (2) Backbone balance,  which implements inter-microbatch load balancing exclusively on the LLM backbone; 
(3) Hybrid balance, which combines interleaved balancing (directly balancing sampled images across ranks) for the encoder with the backbone balance, as described in Fig.~\ref{fig:prog_implementation}. We do not perform intra-microbatch balancing for the LLM backbone.

\noindent
\textbf{Metrics.} 
We evaluate performance using three primary metrics: average iteration time (s), data fetch latency (s), and average loader memory usage per node (GB). For the orchestration evaluation, we also report training throughput in tokens per second (tokens/s). All results are collected over 100 training iterations following an initial warmup period. To ensure a fair comparison, the memory footprint of shadow loaders is excluded from all measurements.

\noindent
\textbf{Testbed.} We conduct our experiments on a training cluster composed of multiple nodes, each equipped with 16$\times$ NVIDIA L20 GPUs (48GB) and 1.8TB of host DRAM. We employ the sidecar mode to launch isolated containers on the host, allocating half of the available CPU cores and memory resources to the resource pool for Ray scheduling. The entire training cluster is interconnected via InfiniBand, with HDFS as the storage backend.


\subsection{Data Preprocessing Architecture Evaluation}

\begin{figure*}[t!]
\centering
\includegraphics[width=\linewidth]{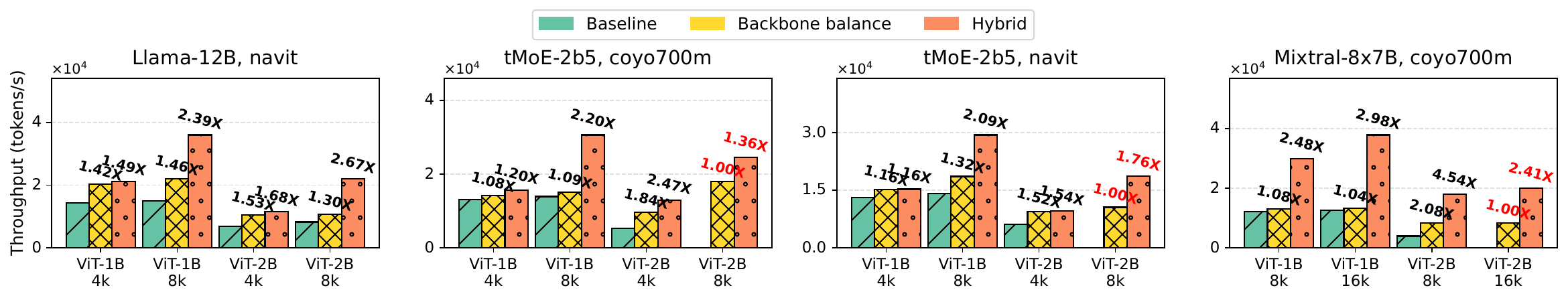}
\vspace{-8mm}
\caption{End-to-end orchestration performance across varying context lengths, dataset groups, and model sizes.}
\vspace{-4mm}
\label{fig:e2e}
\end{figure*}

\begin{figure*}[t!]
    \centering 
    
    \begin{minipage}[t]{0.65\textwidth}
        \centering
        \includegraphics[width=\linewidth, height=4cm]{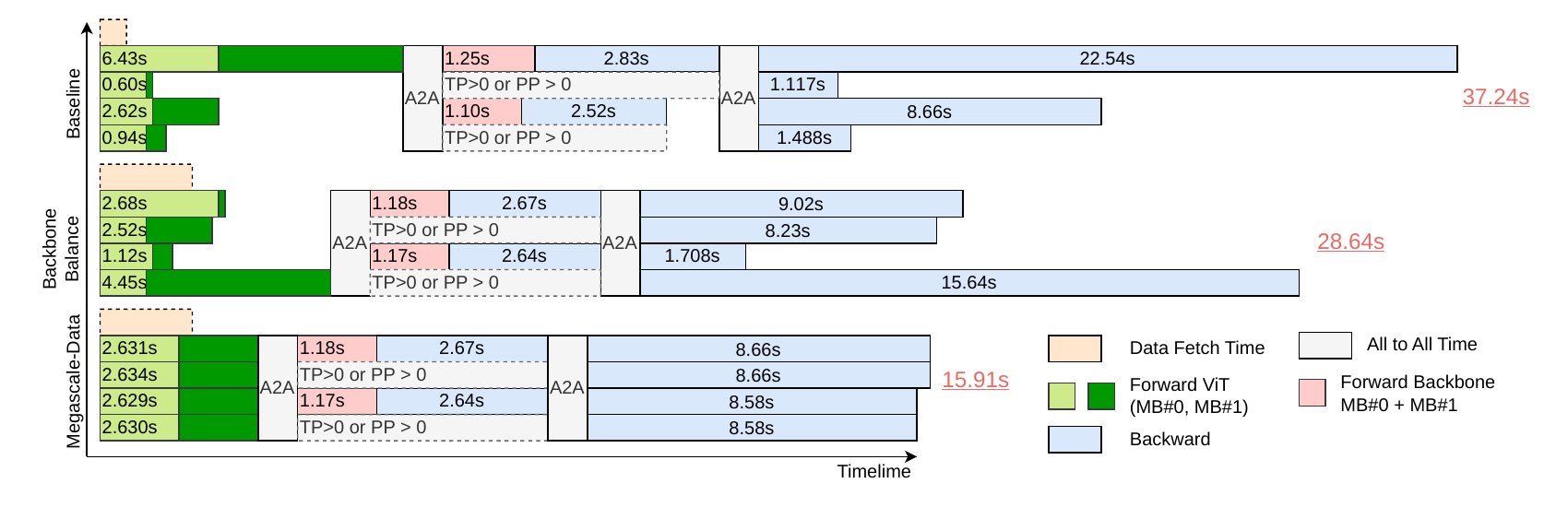}
        \vspace{-8mm} 
        \caption{Case Study for VLM pre-training data orchestration.}
        \label{fig:case_study}
    \end{minipage}%
    \begin{minipage}[t]{0.35\textwidth}
        \centering
        \includegraphics[width=\linewidth, height=4cm]{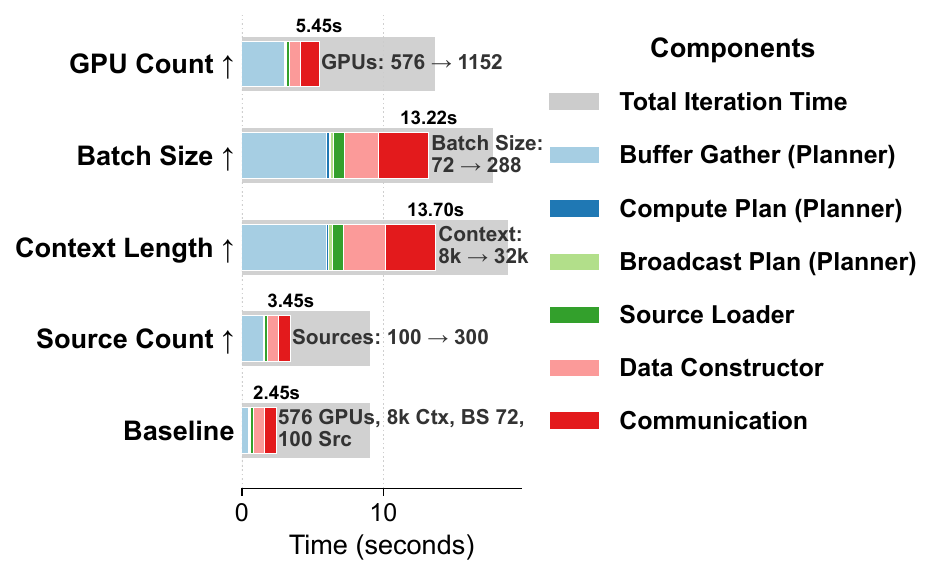} 
        \caption{Time breakdown.}
        \label{fig:perf-breakdown}
    \end{minipage}
\end{figure*}


We demonstrate resource savings achieved by \sys{} using the Llama-12B + ViT-2B combination on 288 and 576 GPUs, with a batch size of 72 for each GPU. To utilize the GPU resources and avoid HBM OOM (out of memory), we size the backbone layer number to fit the model into the GPU memory. We restrict the model layers to 8 and 16, respectively. We measure the memory usage at each node and calculate the average memory per node.

As illustrated in Fig.~\ref{fig:main-loader}, \sys{} achieves up to a $3.63\times$ speedup in training iteration time compared to state-of-the-art loaders, attributable to its efficient load-time data orchestration. It also delivers substantial resource savings, with up to a $13.5\times$ reduction in loader memory usage. This efficiency stems from its ability to eliminate parallelism and source-level data redundancies that existing systems cannot address. While \sys{} incurs a minor fetch overhead due to its coordination mechanisms, this latency is fully masked by the training computation. Notably, the reduced average per-node memory overhead in the 576-GPU configuration is attributed to the auto-partitioning feature, as more abundant CPU resources enable a larger optimization space for the same number of involved data sources.


\subsection{Orchestration Evaluation}

The results of the end-to-end orchestration performance are shown in Fig.~\ref{fig:e2e}. Compared to the non-scheduling baseline, our approach achieves up to 4.54$\times$ (average 1.77$\times$) throughput improvement. The following key observations are made.

\noindent
\textbf{Larger Context Lengths Amplify Heterogeneity.} Longer contexts also introduce greater in-batch heterogeneity. Our balancing strategy capitalizes on this property, yielding higher gains: 4k contexts achieve an average speedup of 1.71$\times$, 8k contexts yield 2.63$\times$, and 16k contexts reach 3.09$\times$. In the absence of load balancing, peak activation memory can induce 
OOM errors, as empirically observed in our ViT-2B experiments. The results also validate the necessity of our load-time balancing strategy, which also mitigates the excessive memory overhead caused by all-to-all collective communication.

\noindent
\textbf{Dataset Characteristics Influence Gains.} Fig.~\ref{fig:sample-distribution} shows that \textit{coyo700m} contains shorter text subsequences than \textit{navit\_data}, leading to greater heterogeneity for image computation with equivalent context lengths. This manifests higher speedups with hybrid balancer: \textit{coyo700m} achieves 2.48$\times$ average (up to 4.54$\times$) versus \textit{navit\_data}'s 2.42$\times$ average (up to 3.47$\times$). For 4k contexts with \textit{navit\_data}, encoder balancing provides limited benefit as small context length results in few samples inside a batch, which offers very limited scheduling space.

\noindent
\textbf{Encoder Scaling Affects Strategy Efficacy.} Hybrid balancing (encoder + backbone) outperforms encoder-only strategies more significantly with larger encoders. For Llama-12B, ViT-2B shows 1.58$\times$ gain versus ViT-1B's 1.41$\times$ under hybrid balancing. This observation is intuitive: the encoder serves as the bottleneck in our current model configuration, and a larger encoder enables more substantial benefits when hybrid balancing is applied.

\subsection{Case studies}\label{sec: case-study}

To demonstrate the benefits of \sys{}, we present a case study using a Llama-12B and ViT-2B model on the \textit{navit\_data} dataset (Fig.~\ref{fig:case_study}). The experiment uses a batch size of 128 with a hybrid parallelism strategy (PP=9, DP=8, CP=2, TP=4), which requires an All-to-All communication collective to transfer features from the ViT encoder to the LLM backbone.

The baseline implementation suffers from a severe workload imbalance in the encoder stage, stemming from variable input image resolutions. This imbalance occurs because the maximum sequence length parameter only constrains the token count for the LLM backbone; it does not regulate the number of visual tokens produced by the encoder, which is dependent on image size. While a naive, microbatch-level balancing scheme is too coarse-grained to resolve this issue, \sys{} integrates a hybrid balancer into the data loading process. This component proactively constructs batches to ensure an even workload distribution across all stages. By addressing the root cause of the imbalance, \sys{} minimizes worker idle time, reducing the end-to-end iteration latency from 37.24s to 15.91s and achieving a 2.34$\times$ speedup.

\subsection{Ablation Studies}\label{sec:ablation}

\begin{figure}[t]
    \centering
    \includegraphics[width=0.65\linewidth]{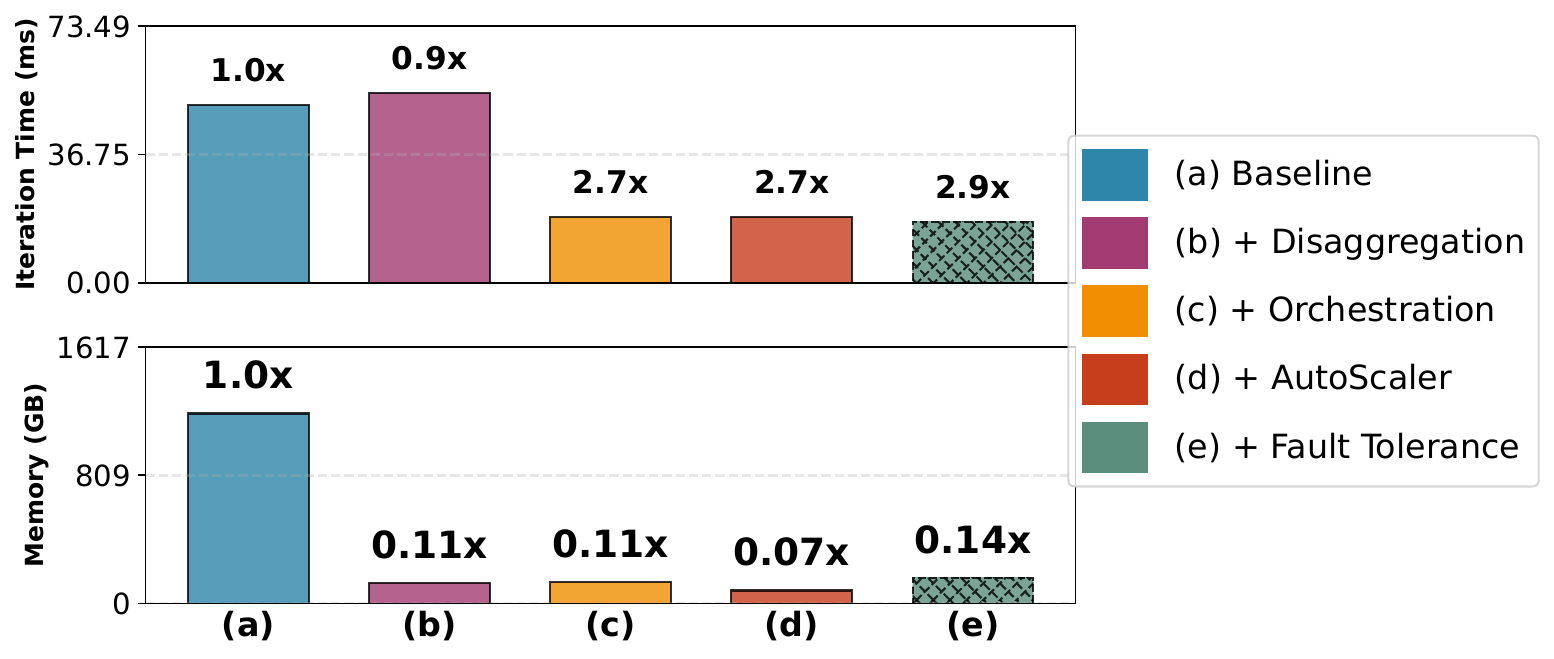}
    \caption{Components Contributions.}
    \label{fig:component-contributions}
\end{figure}

\noindent
\textbf{Time Breakdowns.}
Fig.~\ref{fig:perf-breakdown} illustrates the time breakdown of \sys{} components in training iteration. Across all configurations, including on a 1,152-GPU cluster, the data fetch overhead is minimal and consistently overlapped by the total training iteration time (gray bar).

The overhead scales gracefully with training configuration. Increasing the number of data sources only slightly impacts buffer gathering time. While larger sequence lengths and batch sizes (bounded by GPU HBM) increase overhead from data construction and planning, the training time scales commensurately. This proportional dynamic ensures the data pipeline overhead remains effectively hidden behind computation. Finally, we note that the Planner and Data Constructor are not yet heavily optimized, presenting opportunities for further latency reductions.

\noindent
\textbf{Component Ablations.}
We conducted an ablation study to quantify the contribution of each component in \sys{} (Fig.~\ref{fig:component-contributions}) in the 576-GPU experiment. Disaggregation significantly reduces loader memory usage by eliminating redundant loaders, at the cost of a 10\% latency increase. Building on this, Orchestration achieves a 2.7$\times$ end-to-end performance speedup with negligible memory overhead. The AutoScaler then further optimizes memory utilization by dynamically adjusting resources. Finally, Fault Tolerance, using two shadow loaders, improves the effective training throughput rate (ETTR) by 1.08$\times$ during failures, in exchange for a predictable increase in memory footprint.


\begin{table}[t]
\centering
\caption{API cost for data orchestration under different training setups. Baseline setup is Llama-12B + ViT-2B, 288 GPUs, BS=72, max sequence length (Seq) = 8k. }
\resizebox{\linewidth}{!}{%
\begin{tabular}{lccccc}
\toprule
\textbf{Case} & \textbf{Baseline} & \textbf{+BS 72$\xrightarrow{}$144} & \textbf{+Seq 8K$\xrightarrow{}$16K } & \textbf{+Cluster 288$\xrightarrow{}$1152} & \textbf{+Group 1$\xrightarrow{}$2, 1152 GPUs} \\
\midrule
API:\textit{cost} (s) & 0.004 & 0.006  & 0.012 & 0.107 & 0.106 \\
API:\textit {balance} (s) & 0.016 & 0.027 & 0.173 & 0.357 & 0.195 \\
Iteration Time (s) & 14.31 & 15.98 & 16.91 & 13.56 & 13.56 \\
\bottomrule
\end{tabular}%
}
\label{tab:api-cost}
\end{table}

\noindent
\textbf{API Scalability.}
Table~\ref{tab:api-cost} shows negligible API costs for data orchestration in scaled Llama-12B + ViT-2B experiments on the \textit{navit} dataset. Batch size and sequence length are bounded by GPU HBM capacity. Results confirm the API overhead is highly scalable and much smaller than training iteration time. Although API cost grows with cluster size, setting group size properly effectively controls this increase while maintaining orchestration performance.


\begin{figure}[t]
    \centering
    \newlength{\maxheight}
    \setlength{\maxheight}{4cm}

    \begin{subfigure}[b]{0.47\textwidth}
        \centering
        \includegraphics[height=\maxheight, keepaspectratio]{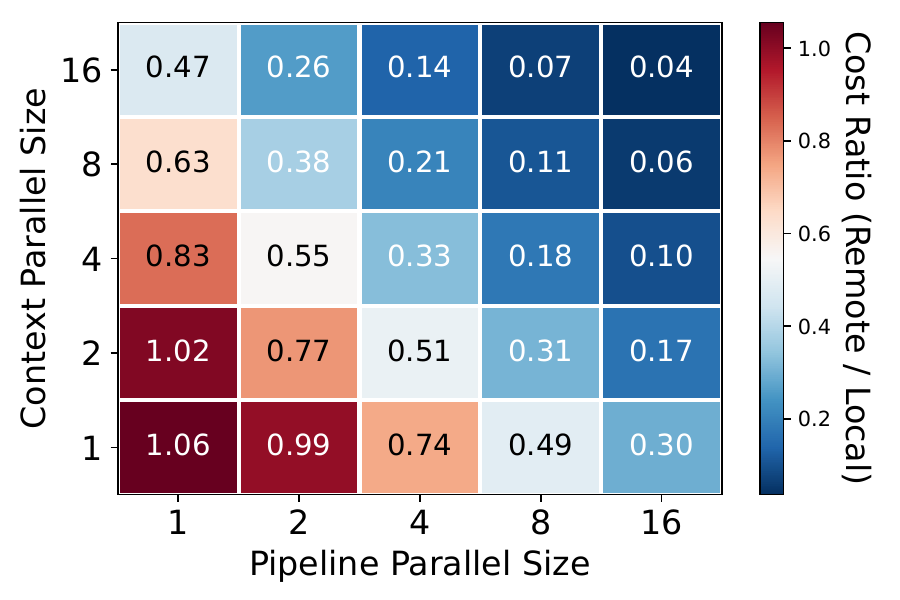}
        \caption{Parallelism redundancy.}
        \label{fig:exp-p-redundancy-removal}
    \end{subfigure}
    \begin{subfigure}[b]{0.47\textwidth}
        \centering
        \includegraphics[height=\maxheight, keepaspectratio]{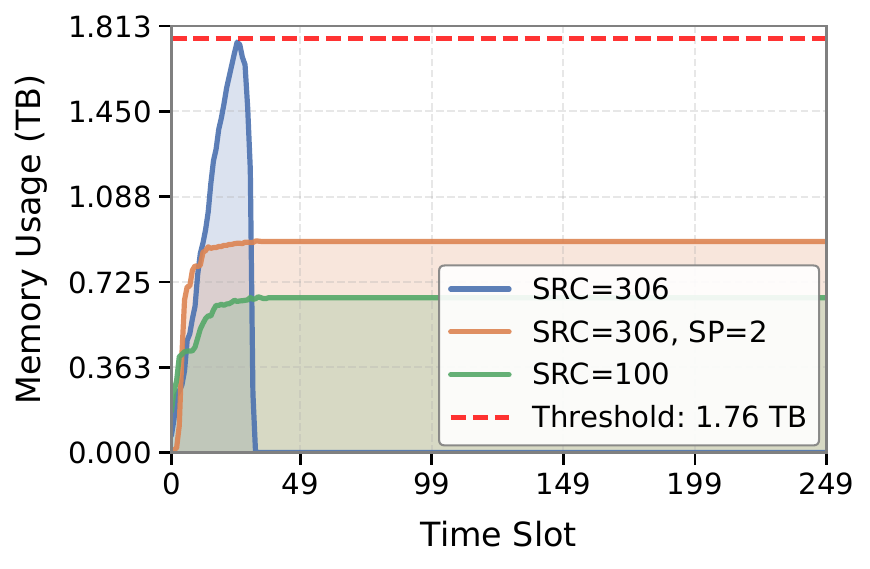}
        \caption{Source redundancy.}
        \label{fig:source_parallel}
    \end{subfigure}

    \vspace{2mm}
    \caption{Peak host memory usage when eliminating (a) parallelism redundancy and (b) source redundancy. SP=2 evenly partitions data sources across data-parallel ranks.}
\end{figure}



\begin{figure}[t]
    \setlength{\maxheight}{4cm}
    \centering
    \begin{subfigure}[b]{0.47\textwidth}
        \centering
        \includegraphics[height=\maxheight, keepaspectratio]{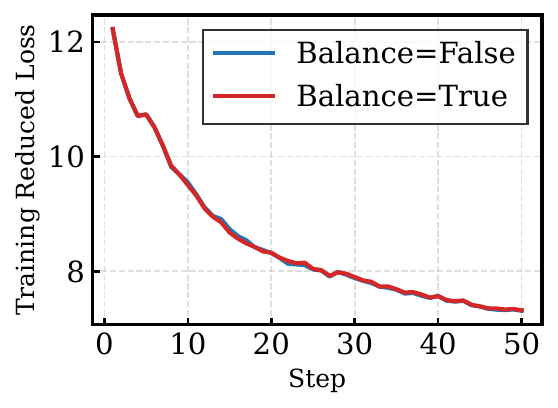}
        \caption{Without CP.}
        \label{fig:non-cp-balance}
    \end{subfigure}
    \begin{subfigure}[b]{0.47\textwidth}
        \centering
        \includegraphics[height=\maxheight, keepaspectratio]{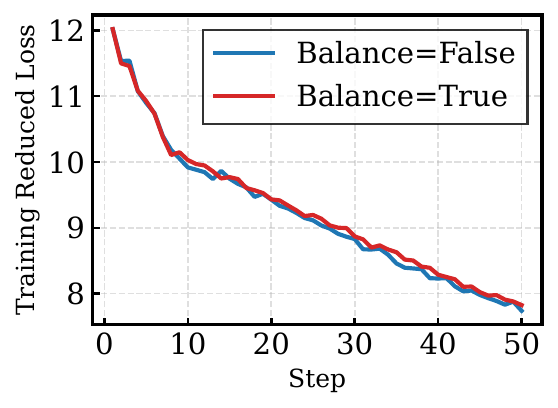}
        \caption{With CP.}
        \label{fig:cp-balance}
    \end{subfigure}
    \caption{Impact of the balancer on training loss convergence. (a) Without Context Parallelism (CP), the balanced loss tightly tracks the baseline. (b) With CP, the balancer introduces minor fluctuations but maintains overall convergence.}
    \label{fig:balance_impact_on_loss}
\end{figure}

\noindent\textbf{Impact on Training Loss.}
Fig.~\ref{fig:balance_impact_on_loss} evaluates the impact of the balancer on training convergence across previous experimented trials: without and with Context Parallelism (CP). As detailed in Sec.~\ref{sec:exp_setup}, we adopt a conservative balancing strategy, where the hybrid balancer performs only inter-microbatch balancing across the backbone while preserving intra-microbatch sample randomness. As shown in Fig.~\ref{fig:non-cp-balance}, in the absence of CP, the training loss of the balanced configuration tightly mirrors the non-balanced baseline. Conversely, when CP is enabled (Fig.~\ref{fig:cp-balance}), the balanced loss exhibits slightly more fluctuation and minor deviations from the baseline, though it still converges effectively. We attribute this variance to the modified sequence partitioning induced by global balancing. This repartitioning alters the distribution of tokens across devices, which inherently introduces minor, acceptable numerical differences during distributed matrix multiplication (GEMM) and summation operations.

\noindent\textbf{Redundancy Removal.}
We evaluate \sys{}'s design efficiency in reducing two primary sources of redundancy: high parallelism and duplicated data sources. First, to quantify parallelism redundancy, we profile memory usage on a simulated backend configured for a large-scale scenario (512 GPUs, BS=512) without source partitioning. As shown in Fig.~\ref{fig:exp-p-redundancy-removal}, memory efficiency gains become significant as context and pipeline parallelism levels increase. We then specifically evaluate the efficacy of source partitioning by isolating the data loader on a cluster configured with TP=16, worker=8 and DP=2. Fig.~\ref{fig:source_parallel} reveals that uniformly partitioning data sources across DP ranks yields substantial reductions in memory overhead, demonstrating our system's comprehensive approach to optimizing memory.

\begin{figure}[t]
\setlength{\maxheight}{4cm}
    \centering
    \includegraphics[height=\maxheight]{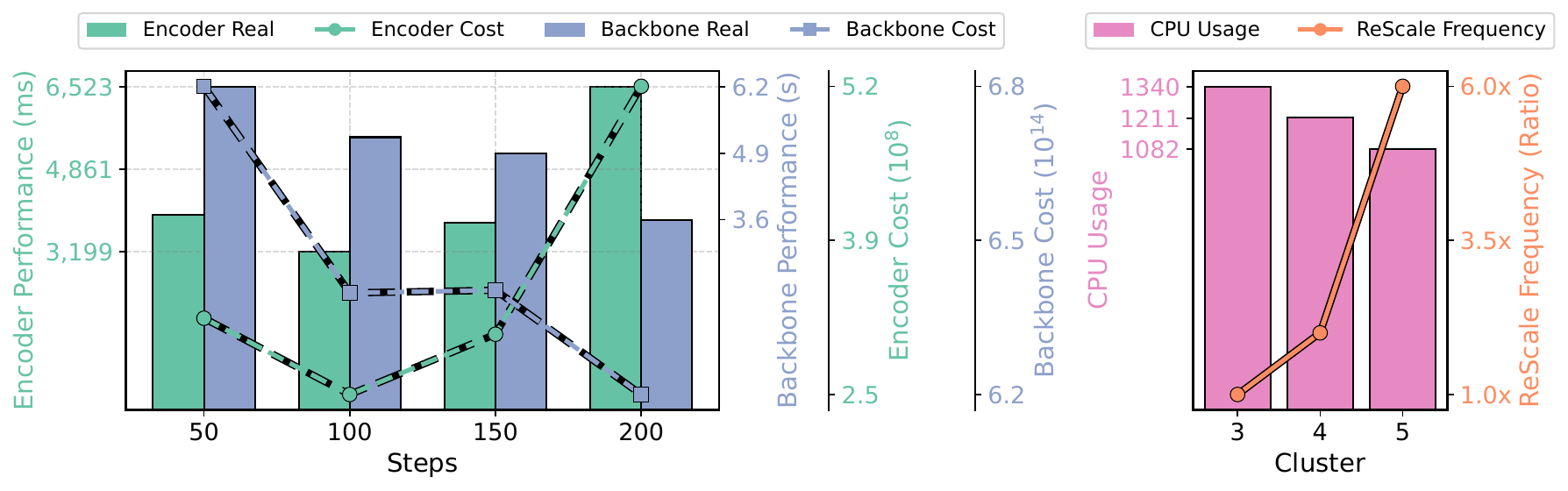}
    \caption{Cost model fidelity and clustering size impact.}
    \label{fig:cost-and-part}
\end{figure}


\noindent\textbf{Cost Model and AutoScaler Analysis.}
Fig.~\ref{fig:cost-and-part} validates our performance model and evaluates the trade-offs of auto-partitioning. The left panel validates our model's accuracy, showing that its predictions for the encoder and a single-layer backbone closely track empirical measurements. The right panel illustrates the trade-off between improved load balancing from more partitions and the resulting increase in rescaling costs. We identify a partition size of 4 as the optimal balance for the evaluated production workloads.

\begin{figure}[t]
\setlength{\maxheight}{4cm}
\centering
\includegraphics[height=\maxheight]{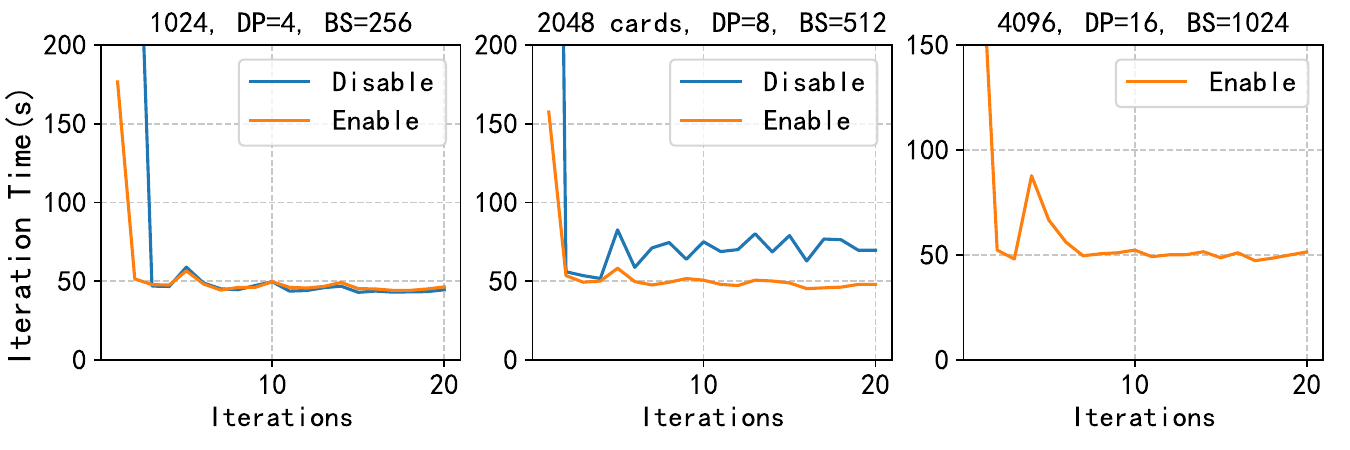}
\caption{Scalability Advantages of the Actor Model.}
\label{fig:sale-eval}
\end{figure}

\noindent
\textbf{Scalability Advantages of the Actor Model}
To evaluate the scalability of our disaggregated structure (Fig.~\ref{fig:sale-eval}), we train a pure-text model, comparing \sys{} against a direct-transfer baseline that bypasses the Data Constructor. While performance is comparable at 1k GPUs, scaling to 2k GPUs reveals severe connection overhead in the baseline, which suffers a 10$\times$ data fetch latency increase compared to \sys{}. At 4k GPUs, the baseline completely collapses due to communication bottlenecks. In contrast, \sys{} sustains throughput via data redistribution, underscoring the Data Constructor's critical role in extreme-scale training.

\section{Related Work}\label{app:related-work}

\noindent
\textbf{Remote dataloader.}
Disaggregated data loading architectures~\cite{DPP, tf_data_service} offload data preprocessing to remote workers to improve resource management and prevent data stalls. Subsequent systems introduced optimizations such as auto-caching~\cite{graur2022cachew}, auto-scaling, selective worker placement~\cite{pecan}, and combinatorial pipeline search~\cite{zhao2024cedar}. However, these approaches are misaligned with LFM training, where single-epoch processing renders data heterogeneity and memory redundancy from numerous sources—not CPU limitations—is the primary bottleneck. In contrast, \sys{} is specifically designed to eliminate this memory redundancy and provide the flexible, multisource orchestration required by LFMs.



\noindent
\textbf{Multimodal Training Frameworks.}
Recent frameworks seek to boost the efficiency of multimodal training by addressing model and data heterogeneity~\cite{DistMM, optimus, DistTrain}. These systems primarily employ model-level optimizations, such as applying distinct parallelism to submodules~\cite{DistMM}, scheduling computations to fill pipeline bubbles~\cite{optimus}, and disaggregating model components for independent resource management~\cite{DistTrain}. While effective at balancing the model's computational workload, these methods overlook opportunities in data orchestration. In contrast, \sys{} introduces dynamic, load-time balancing through a disaggregated data architecture, optimizing performance without adding communication overhead to the critical path or increasing memory consumption.

\noindent
\textbf{Fault-tolerant training.}
Previous studies~\cite{robust-tpu, robust-shlab, megascale, aegis, byterobust} have introduced robust AI infrastructures to automatically address various incidents during training.
These systems employ a variety of techniques for failure detection and diagnosis~\cite{robust-shlab, megascale, aegis, mycroft}, failover procedure optimizations~\cite{pathway, byterobust}, and checkpoint recovery~\cite{checkfreq, check-n-run, gemini-aws, wan2024bytecheckpointunifiedcheckpointingllm}.
Unlike prior work, \sys{} enhances the fault tolerance of the data loading module by providing differential checkpointing for its components and utilizing shadow loaders for hot recovery to ensure consistent data delivery during training. 
\section{Future Work}
\noindent
\textbf{Replay Mode.} While our system supports dynamic data mixing to adjust sampling ratios at runtime, many production training workloads employ predictable learning schedules. For such cases, the orchestration plan for each step can be pre-computed offline. By decoupling planning from execution, the runtime can transition to a "Replay Mode" that executes pre-computed schedules, significantly reducing online Planner overhead. This architectural shift allows the Planner to prioritize high-level health monitoring and fault tolerance over per-step data orchestration.

\noindent
\textbf{Ahead-of-Fetch Load Balancing.} Current source loaders perform load balancing reactively after fetching data from the storage. We aim to explore Ahead-of-Fetch balancing by leveraging early metadata acquisition (e.g., pre-computing sample costs and embedding them directly into the storage). This design simultaneously enhances architectural flexibility and system efficiency: by decoupling metadata retrieval from the source loader, it allows for more flexible transformation placement within the data preprocessing pipeline and creates potential cost savings.

\noindent
\textbf{Strategy Optimizer.}
Since our data plane is declarative, we plan to develop an optimizer that automatically rewrites and fuses orchestration strategies. Specifically, we aim to explore graph rewriting techniques to fuse orchestration primitives and translate the execution plan into efficient, low-latency C kernels.

\section{Conclusion}
\sys{} tackles the challenge of scalable multi-source data preprocessing and orchestration for large foundation model training. The actor-model preprocessing framework eradicates source and parallelism redundancy in multisource data loading, ensuring scalable preprocessing for LFM training jobs. The declarative data plane simplifies the programming of intricate cross-module, multi-source data scheduling and mixture sampling strategies. Leveraging source autoscaling, we partition and organize heterogeneous sources effectively.  Experiments show that \sys{} outperforms state-of-the-art baselines significantly in terms of throughput and resource efficiency, while maintaining scalability and robustness.


\clearpage

\bibliographystyle{plainnat}
\bibliography{paper}

\clearpage



\end{document}